# Lanthanide Ion Electronic Structure Controls Magnetic Excitations in Topological Quantum Ferrimagnets $LnMn_6Sn_6$ (Ln = Tb, Dy, Ho)

Kelsey A. Collins,[1,2*] Jacob Pfund,[1,3] Michael R. Page,[1] Menka Jain,[3] Michael A. Susner,[1] and Michael J. Newburger[1]

[1]Materials and Manufacturing Directorate, Air Force Research Laboratory, Wright-Patterson Air Force Base, OH 45433, USA

[2]Core4ce, Dayton, OH 45433, USA

[3]Department of Physics and Institute of Materials Science, University of Connecticut, Storrs, CT 06269, USA

*corresponding author: kelsey.collins.1.ctr@afrl.af.mil

## Abstract

The $LnMn_6Sn_6$ family of topological magnets is a promising platform for next-generation spintronic and magnonic technologies. However, the influence of the lanthanide ion ($Ln^{3+}$) on the excited-state spin dynamics, or magnons, remains a critical knowledge gap. Here, we present the first comparative study of the magnetic dynamics in $LnMn_6Sn_6$ materials (Ln = Tb, Dy, Ho) using Brillouin light scattering. Our findings reveal a direct correlation between the lanthanide ion's intrinsic properties and the magnon behavior. We demonstrate that the magnon frequency in the absence of an applied magnetic field is primarily dictated by the strength of the lanthanide exchange coupling, as modeled by its relationship with the de Gennes factor. The response of the magnon to an applied field is influenced by material's gyromagnetic ratio and the overall anisotropy of the material, which are dictated by total angular momentum and the anisotropy of the lanthanide sublattice, respectively. These results establish that simple lanthanide substitution provides a powerful and predictable method for tuning magnon properties, enabling the rational design of materials for advanced technological applications.

# A. Main text

## 1. Introduction

Topological magnets, which exhibit intrinsic coupling between spontaneous magnetization and emergent electronic topology, provide opportunities to discover novel condensed-matter phenomena and to develop functional platforms for quantum technologies. In such materials, nontrivial band topology generates robust boundary states and exotic electronic states such as Dirac and Weyl semi-metals.[1–11] Magnetic ordering in topological materials breaks time-reversal symmetry, enabling effects such as the quantum anomalous Hall effect and topological magneto-electric effects[5–7,9,10] that are not accessible in nonmagnetic topological insulators.[12]

This interplay of magnetism and topology further enables the emergence of novel quasiparticles and excitations,[11] including skyrmions[8,10] and topological magnons[3,5] - opening up avenues for exploring fundamental condensed matter concepts and paving the way for new technological applications. Topological magnons are of particular interest because their protected edge and surface modes provide a promising foundation for magnonic information transport and ultra-precise sensing. The boundary localization and symmetry protection of these modes render their spectra and propagation highly sensitive to perturbations, including structural distortions, strain, and interfacial modifications.

One promising family of topological magnets is the $LnMn_6Sn_6$ family, where Ln is a trivalent lanthanide or rare earth cation.[13–16] Although the impact of the trivalent lanthanide ion on the magnetic ground state of the $LnMn_6Sn_6$ family is well established,[13–16] its influence on excited-state spin dynamics and magnonic excitations remains insufficiently understood, thus leaving a critical gap in our technological understanding of these materials.

We seek to determine how the lanthanide cation affects the magnetic excitations of these materials, including the magnon frequency and lifetime, and whether this dependence can be leveraged to enable specific applications. Understanding these excitations is of particular interest due to the nontrivial topology inherent to the $LnMn_6Sn_6$ family, which positions it as a promising potential platform for hosting topological magnons and quasiparticle excitations. Such excitations are of fundamental and technological interest. From a fundamental standpoint, the interplay of topological electronic structure and magnetism can give rise to dispersionless magnon bands and protected edge modes.[3] Technologically, topological magnons have potential applications in spintronics, where such spin currents may enable advanced computing architectures, and in high-frequency electronics, where surface-confined magnons could enhance frequency resolution and operational bandwidth.

In the $LnMn_6Sn_6$ family, the topological magnetism arises from the Mn atoms which are arrayed into two-dimensional kagome lattices comprised of corner sharing equilateral triangles in the crystallographic *ab* plane (**Figure 1**a and b). This lattice topology has been shown to give rise to exotic phenomena in magnetic materials,[17–19] prompting extensive interest in the electronic and magnetic behavior of this materials family. Within each Mn layer, the Mn spins are coupled to one another ferromagnetically, i.e. the electron spins are aligned in the same direction (Figure 1c). These Mn-containing kagome layers are separated by a complex arrangement of two crystallographically distinct Sn atomic sites, one of which is located at the center of the trigonal prism bounded by the nearest six Mn atoms and which mediates ferromagnetic superexchange coupling between neighboring 2D Mn layers. The remaining Sn crystallographic site is located in the center of the trigonal planes defined by the $Ln^{3+}$ cations, which sit in a hexagonal coordination geometry. (Figure 1a and c).

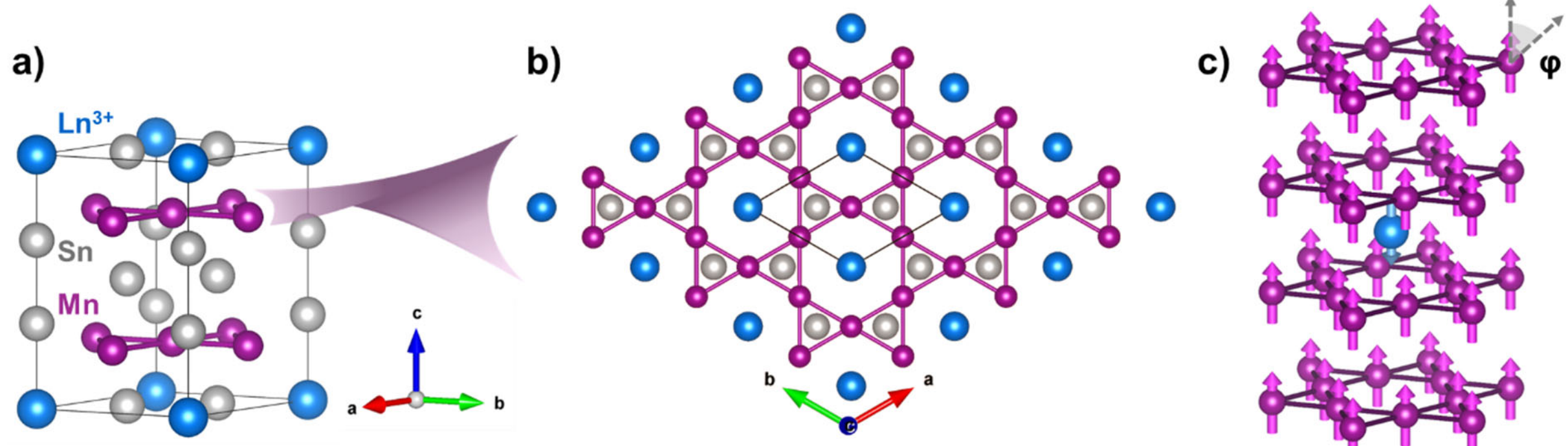

Figure 1: Crystallographic structure of $LnMn_6Sn_6$; in all subsets, Mn, $Ln^{3+}$, and Sn are represented by purple, blue, and gray spheres, respectively. a) Unit cell. b) View down the crystallographic *c* axis, showing the Mn-based kagome lattice that gives rise to topological magnetic behavior. c) Magnetic structure showing the relative orientation of the magnetization of the different sublattices. For $TbMn_6Sn_6$, the magnetization at $T < 310$ K is as-depicted, along the *c* axis, but for $DyMn_6Sn_6$ and $HoMn_6Sn_6$ the magnetization is oriented at a characteristic angle, φ, away from the *c* axis (φ = 45° for $DyMn_6Sn_6$ and 49° for $HoMn_6Sn_6$).

The identity of the $Ln^{3+}$ ion influences the magnetic ground state in multiple ways.[13] For example, when the $Ln^{3+}$ ion is $Gd^{3+}$ - $Er^{3+}$, the coupling between it and the Mn-based kagome layer is antiferromagnetic, giving rise to a net ferrimagnetic ground state.[15] The *magnitude* of this coupling between the $Ln^{3+}$ and Mn-based kagome layer thus determines the magnetic ordering temperature, $T_C$, of the material. Additionally, the crystal field splitting-induced *single ion anisotropy* of the $Ln^{3+}$ ion dictates the preferred orientation of the magnetization of the whole material.[13,20] In the absence of an anisotropic paramagnetic ion, the anisotropy of the Mn-based kagome layers controls the preferred orientation of the magnetization, resulting in the material having magnetization where the easy direction is within the crystallographic *ab* plane. For example, in $GdMn_6Sn_6$, the $Gd^{3+}$ ion is *fully isotropic* due to its $4f^7$ electron configuration and has no anisotropy with which to bias the magnetization, thus resulting in $GdMn_6Sn_6$ evoking so-called easy-plane behavior where the easy axis lies along the *ab* plane with negligible anisotropy within the basal plane (here, the *a* axis was found to be slightly easier).[21] In the case of an *anisotropic* $Ln^{3+}$, such as $Tb^{3+}$, the anisotropy of the $Tb^{3+}$ ion forces the magnetization to preferentially orient along the crystallographic *c* axis, resulting in a *c*-aligned easy axis orthogonal to the kagome layers (Figure 1c).[22] The combination of spin-orbit-coupling with this magnetic structure gives rise to topological magnetic skyrmions in $TbMn_6Sn_6$.[16,23] In contrast, *moderately anisotropic* ions such as $Dy^{3+}$ and $Ho^{3+}$ give rise to magnetic ground states in which the magnetization is neither easy axis or easy plane, but is instead easy cone, oriented at a characteristic angle, φ, away from the *c* axis with φ (Dy) = 45° and φ (Ho) = 49° (Figure 1c).[13,15]

The same interplay between lanthanide cation anisotropy and the anisotropy of the kagome structured Mn planes that yields these different magnetic structures also gives rise to an unusual spin reorientation transition. In $TbMn_6Sn_6$, as established above, the magnetic ground state is a collinear easy-axis ferrimagnet. However, at $T_{SR}$ = 312 K (compared to a $T_C$ of 450 K), the anisotropy of the $Tb^{3+}$ decays such that the magnetic structure undergoes a spin-reorientation from alignment along *c* to in-plane with *ab* plane.[24]. Because the anisotropy of the lanthanide ion stems from 4*f* electrons and orbitals, which only weakly interact with the crystal field, both the magnetic moment and the anisotropy of the lanthanide sublattice decays gradually with temperature.[25] In contrast, the magnetic moment and anisotropy of the Mn sublattice is relatively constant with temperature.[25] This results in a cross-over point, at which the anisotropy of the Mn sublattice is greater than that of the lanthanide sublattice and the magnetization goes from easy-axis to easy-plane. This spin reorientation also occurs in $DyMn_6Sn_6$ and $HoMn_6Sn_6$; however, due to the lower anisotropies of $Dy^{3+}$ and $Ho^{3+}$ compared to that of $Tb^{3+}$, the transition occurs at lower temperatures than in $TbMn_6Sn_6$.

As mentioned above, while the influence of the trivalent lanthanide ion on the magnetic ground state of the $LnMn_6Sn_6$ family is well established both theoretically and experimentally, far less attention has been dedicated to its effect on magnetic excited states, specifically spin wave excitations, or magnons. To our knowledge, there has been only one report on the spin wave modes in this material family.[26] That work examined the ferromagnetic resonance modes in $MgMn_6Sn_6$ and suggested that orbital angular momentum of the Mn contributed significantly to the magnetic anisotropy and nontrivial electronic structure of the compound. However, this material contains a diamagnetic cation ($Mg^{2+}$) and thus could not reveal the influence of the lanthanide ion on the magnetic excitation spectrum.

Herein, we investigate the magnon behavior of three members of this family containing paramagnetic lanthanide ions ($TbMn_6Sn_6$ (**1-Tb**), $DyMn_6Sn_6$ (**2-Dy**), and $HoMn_6Sn_6$ (**3-Ho**)) using Brillouin light scattering, which enables probes magnons in the GHz frequency regime. Our aim is to understand the influence of the lanthanide ion on the magnon frequency and lifetime, as well as the effect of an external magnetic field and temperature on the magnon frequency. Our results reveal that the frequency of magnons is strongly dependent on the identity of the lanthanide ion. Quantitative fitting of the magnon frequencies for these three materials reveals two key trends: 1) the single-ion anisotropy of the lanthanide ion controls the magnon frequency, and 2) the total angular momentum of the lanthanide ion governs the material's gyromagnetic ratio, which dictates how the magnon frequency changes under an applied magnetic field. These results indicate that facile frequency tuning of the magnon can be achieved by substitution of the lanthanide ion, which positions this materials family as a highly promising platform for future magnonic and spintronic technologies.

## 2. *Results and discussion*

In Brillouin light scattering (BLS), coherent light interacting with a material has a probability to undergo inelastic scattering with a quasiparticle, producing a small Stokes or Anti-Stokes shift in the scattered photon. This process is similar to Raman spectroscopy; however, BLS is operative at gigahertz to megahertz frequencies, enabling measurement of low energy excitations such as acoustic phonons and ferromagnetic magnons.[27–30] In this work, we utilize micro-BLS in a backscattering geometry such that the light is reflected off the sample and the excitation and collection path are co-linear and performed with the same objective. We performed BLS measurements on flux-grown single crystals of **1-Tb** (Figure 2a), **2-Dy**, and **3-Ho** with incident laser light of 532 nm. The typical laser power at the sample was 8mW with a focused spot size of ~3μm. Further, the incident and scattered light is normal to the crystallographic *ab* plane (parallel to the *c*-axis) of the three samples (Figure 2b and c).

The **1-Tb** material hosts a single Stokes and anti-Stokes magnon that increases monotonically in frequency with a magnetic field applied parallel to the *ab* plane of the crystal. At zero applied magnetic field and at room temperature, this magnon has a frequency of approximately |17.6| GHz (Figure 2d). A Lorentzian fit of these peaks yields an exact Stokes frequency ($f_S$) of −17.43 ± 0.13 GHz and Anti-Stokes frequency ($f_{AS}$) of 17.99 ± 0.13 GHz, respectively. This slight difference in the absolute frequencies of the Stokes and Anti-Stokes peaks represents a miniscule absolute energy (0.5 GHz ≈ 0.2 K) that we do not consider meaningfully significant, but may stem from the Dzyaloshinskii-Moriya interaction, which can manifest as slight asymmetries in the Stokes and Anti-Stokes absolute frequencies in BLS.[31,32] The magnon reaches a frequency of |27.9| GHz at our highest applied field of ~200 mT (Figure 2d).

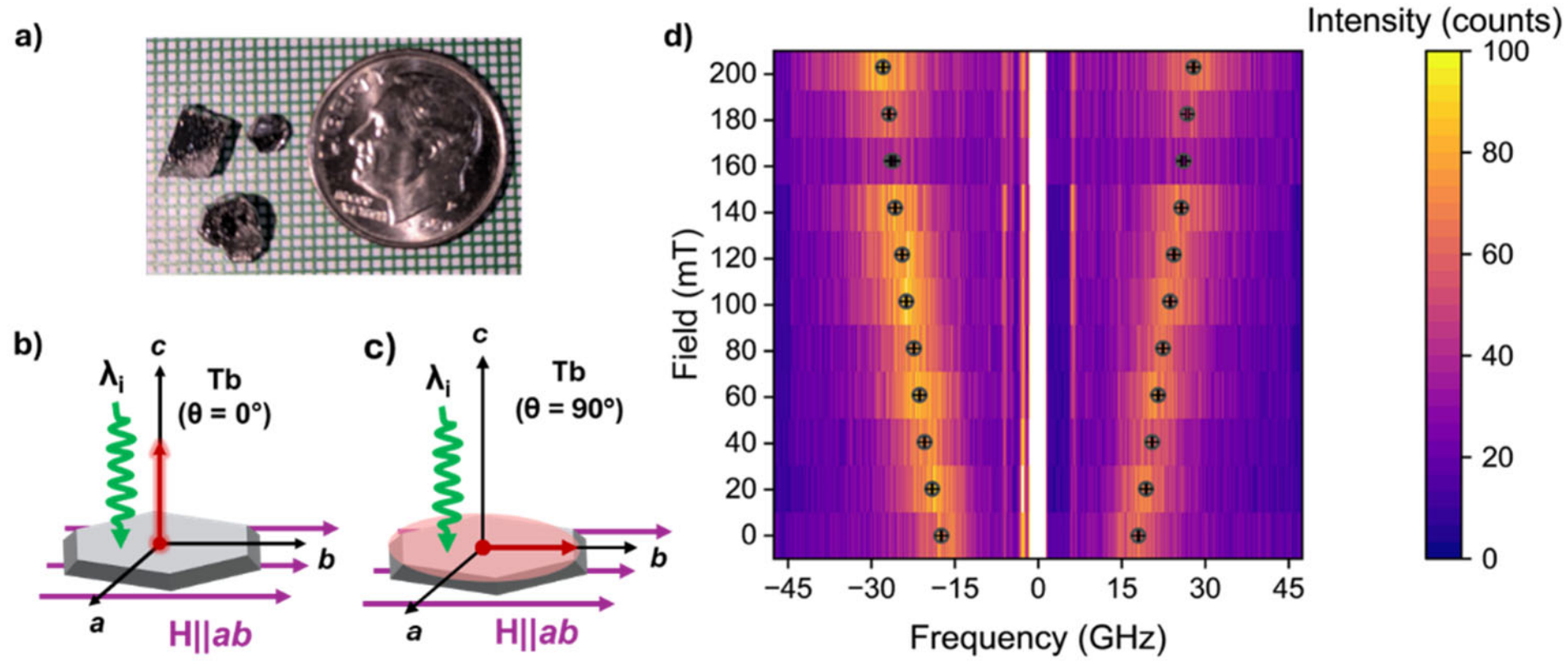

Figure 2: a) Optical photograph of crystals of **1-Tb** used in this work. b-c) Schematic shows the orientation of the magnetization (red arrow) of $TbMn_6Sn_6$ crystal (gray hexagon) when axial (b) and planar (c) and the experimental configuration of the incident laser, $\lambda_i$, (green wavy arrow) and the external magnetic field, H, (purple arrows) applied along the crystallographic *ab* plane. d) Heatmap of field swept BLS spectra of $TbMn_6Sn_6$ with fitted magnon frequencies overlaid as gray points.

This resembles the behavior expected for a ferrimagnet when the applied magnetic field is parallel to the material's magnetization according to the Kittel equation for a hexagonally symmetric material:

$$\text{eq. (1)} \quad f_m = \frac{\gamma}{2\pi}\sqrt{(H_{||} + H_{A,in} + H_{ex})(H_{||} + H_{ex} + H_{A,out} + 4\pi M_S)}$$

where $f_m$ is the frequency of the magnon, γ is the gyromagnetic ratio (Hz/Oe), $H_{||}$ is the applied magnetic field parallel to the *ab* plane of the material (Oe), $H_{A,in}$ is the in-plane anisotropy field (Oe), $H_{A,out}$ is the out-of-plane anisotropy field, $H_{ex}$ is the exchange field, and $4\pi M_S$ is the saturation magnetization value (Oe).[28,33–35]

However, this behavior is unexpected as the magnetic field is applied orthogonal to the magnetization easy axis (Figure 2b). The applied magnetic field range is lower than the saturation field (i.e. the field required to fully re-orient the magnetization from axial to in-plane (Figure 2b)) as measured by vibrating sample magnetometry (Figure S-2), and therefore should not be strong enough to reorient the magnetization in-plane.[33] When the magnetic field is applied orthogonal to the magnetization, the magnon frequency decreases until the magnetic field reaches the saturation field, and at applied fields higher than the saturation field, the magnon frequency linearly increases.[33]

Thus, we hypothesize that the incident laser is inducing the spin reorientation transition, such that the magnetization in 1-**Tb** is planar and not axial under BLS conditions. The laser could be inducing this either via local heating or by nonthermal magneto-optical effects.[36–38] As the spin reorientation temperature is modest in 1-**Tb** (312 K (39 °C)), this transition is readily accessible (Figure S-1). Furthermore, the field dependence of the room temperature **1-Tb** magnon spectra matches the field dependence of **2-Dy** and **3-Ho** (*vide infra*), which are both expected in the fully planar orientation at room temperature, as they host $T_{SR}$'s of 272 K and 185 K, respectively (Figure S-1).

In order to confirm the spin-reorientation as a mechanism for the observed frequency dependence, variable temperature BLS experiments were performed at zero applied magnetic field from 253 K (–30 °C) to 373 K (100 °C) (Figure 3a). BLS is sensitive to structural and magnetic phase transitions, so we would expect a discontinuity in the magnon frequency or linewidth in the magnon spectrum at $T_{SR}$.[27,30,33,36,39–41] Instead, we only observe a gradual decrease in the magnon frequency with temperature, due to thermal population of magnetic excited states and a corresponding decrease in the magnetization saturation value (Figure 3b).[33,42,43] Similarly, the linewidth of the magnon peak does not change appreciably with temperature (until 263 K (–20 °C) when the peaks are broad and weak in intensity) (Figure 3c). Below room temperature, the frequency of the magnon increases slightly, as expected given that the saturation magnetization value also increases at lower temperatures (Figure 3b),[33,42,43] however, this change is accompanied by a concomitant decrease in magnon intensity. In fact, at only 253 K (–30 °C), the magnon peaks are no longer observable at all. This disappearance of a magnon upon cooling is highly unusual, as typically we would expect a magnon to disappear upon a transition to an unordered magnetic state at temperatures above $T_C$, not below it.[44]

We hypothesize that this disappearance of the magnon signal is possibly induced by either (1) the spin reorientation transition back to the easy axis state or (2) that it is an optical rather than a magnetic phenomenon. In opaque materials such as the metals probed here, the relative width of the peak is related to the ratio of the imaginary and real parts of the index of refraction.[45] If the imaginary component of the index of refraction changes sufficiently with temperature, the peak would broaden out and no longer be observable. The exact cause of the magnon signal disappearance is beyond the scope of this study and merits deeper future investigation.

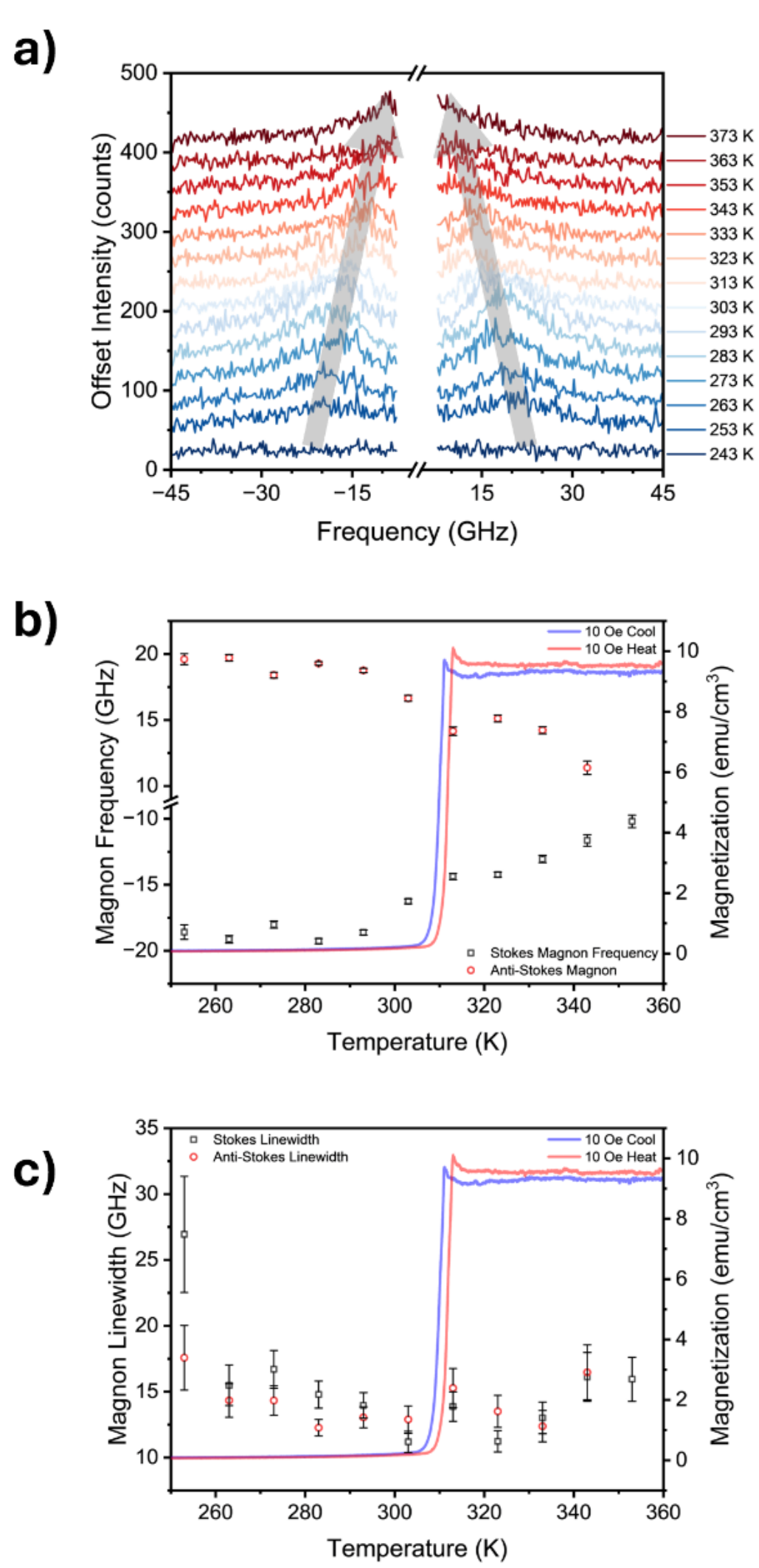

Figure 3: a) Variable temperature BLS spectra of $TbMn_6Sn_6$ at zero applied field; fitted magnon frequency (b), and linewidth (c) with temperature. Gray arrows in (a) are guides to the eye to highlight the shift in peak frequency with temperature. Blue and red lines in (b) and (c) are the zero field cooled and field cooled magnetization from VSM, respectively, showing the lack of a phase transition in the BLS data at $T_{SR}$.

To better understand the frequency dependence of the Tb-based compounds, we compare their behavior to the easy-cone materials **2-Dy** and **3-Ho**. While all 3 compounds show similar monotonic behavior with the absolute frequency increasing with the applied field, the **2-Dy** and **3-Ho** materials host magnons at much lower frequencies. At zero applied magnetic field, the magnon in **2-Dy** has an approximate frequency of 9 GHz ($f_S$ = -9.4 ± 0.7 GHz and $f_{AS}$ = 8.5 ± 0.6 GHz (Figure S-16)), which increases to a maximum of 17.5 GHz ($f_S$ = -16.7 ± 0.2 GHz and $f_{AS}$ = 18.3 ± 0.2 GHz at 200 mT (Figure 4a, Figure S-27)). The magnon in **3-Ho** has a frequency of approximately 5 GHz ($f_S$ = -4.5 ± 0.4 GHz and $f_{AS}$ = 5.6 ± 0.1 GHz (Figure 4a, Figure S-28)) at zero applied field, slightly lower than that of the **2-Dy**. However, upon the application of a nonzero magnetic field, the frequency of the magnon increases to 18.4 GHz ($f_s$ = -18.1 ± 0.2 GHz and $f_{AS}$

= 18.7 ± 0.2 GHz) similar to **2-Dy** (Figure 4a, Figure S-38, and Figure S-39). Consistent with the magnon observed in **1-Tb**, the magnons in **2-Dy** and **3-Ho** are very broad at all applied magnetic fields (Figure S-40).

In our discussion, we focus on two notable aspects of the BLS peak shape, namely the width and center, which provide information about magnons in the materials. The broadness of the magnon peaks in **1-Tb**, **2-Dy**, and **3-Ho** is notable, with a full-width half maximum of approximately 11, 13, and 11 GHz, respectively (Figure S-40). These large linewidths are possibly indicative of short magnon lifetimes.[45–47] Such broad magnon peaks would be prohibitively difficult to observe using ferromagnetic resonance spectroscopy, highlighting the value of BLS experiments for measuring magnons with such short lifetimes or non-zero wavevectors. Magnon relaxation can occur by multiple different scattering processes including those between magnons and other magnons; magnons and phonons; or magnons and electrons (in metallic magnets such as these). Magnon relaxation can also be induced by scattering due to crystallographic defects. As the materials we study here are bulk crystals, we do expect defect broadening to contribute to the magnon linewidth. Despite the many potential mechanisms of magnon relaxation in these materials, we are able to resolve their unique magnon signatures. Additionally, as discussed in the interpretation of the variable temperature data, the opacity of the materials could also increase the observed linewidths. Since all three materials are opaque metals, they have a large imaginary component of the optical index of refraction which leads to a larger linewidth in the observed peaks relative to translucent materials.[45]

The magnon frequency, on the other hand, is dependent upon three major material specific parameters referenced above: 1) the material's saturation magnetization, 2) the magnetic exchange stiffness, and 3) the anisotropy (eq. 1).[28,33–35] The saturation magnetization values of the three materials should be similar, as the $Tb^{3+}$, $Dy^{3+}$, and $Ho^{3+}$ ions differ in electron count by one *f* electron each, and their effective magnetic moments ($\mu_{eff} = g_J(J(J + 1))^{1/2}$) as free ions are very similar (9.7 $\mu_B$ for $Tb^{3+}$, 10.6 $\mu_B$ for $Dy^{3+}$, and 10.6 $\mu_B$ for $Ho^{3+}$).[48,49] Furthermore, the Mn-based kagome layers are the source of most of the magnetic moment in the material.[24,50] Therefore, the saturation magnetization should be largely unchanged by the identity of the lanthanide. Indeed, when all three materials are measured by VSM, they display near identical saturation magnetization values of ~210 emu/cm$^3$ (Figure S-2 and Figure S-3). Thus, we do not expect the saturation magnetization to be the source of significant differences in magnon frequency, and in our analysis this value is fixed to 210 emu/cm$^3$ for all three materials.

The simulated frequency dependence using a modified version of equation 1 for all three materials is shown in Figure 4. The large non-zero magnon frequencies in the absence of an applied magnetic field are due to the in-plane anisotropy and the exchange fields. The in-plane anisotropy field, $H_{A,in}$, stems from a small sixth order anisotropy term and is consequently smaller than the out-of-plane anisotropy fields by several orders of magnitude. This makes it difficult to experimentally observe in BLS. This in-plane hexagonal anisotropy is further obscured by a modest magnetic shape anisotropy that arises from the physical shape of the crystals as thin platelets. This effect can be observed in the in-plane rotation dependence of the zero field magnon in **1-Tb**, which has a uniaxial symmetry (ESI, Figure S-15). If the intrinsic hexagonal in-plane magnetocrystalline anisotropy was stronger than the shape anisotropy, the zero field magnon would instead display a six-fold symmetry. Therefore, we treat **1-Tb**, **2-Dy**, and **3-Ho** as having negligible intrinsic in-plane anisotropy fields in the planar spin configuration.

Consequently, the parameter that most strongly affects the zero-field magnon frequency is the exchange field. Generally, the influence of the exchange field on the magnon frequency is small, as the exchange field is dependent on the momentum of the sampled magnon, $H_{ex} = Dq^2$, where $D$ is the exchange stiffness coefficient and $q$ is the wavevector of the sampled magnon. Unlike in FMR, where $q$ is strictly zero, $q$ is of intermediate value in BLS ($q \approx 10^7$ m$^{-1}$), resulting in the exchange field being on the order of $10^1$ - $10^2$ Oe. This has a minimal effect on the magnon frequency, *except* at zero field, where it enables a large nonzero magnon frequency.[34,35,51] We expect the exchange stiffness coefficient ($D$) to decrease from **1-Tb** to **2-Dy** to **3-Ho**, following the trend of their respective ordering temperatures. The ordering temperature $T_C$ of the three materials is a suitable proxy for $D$ as they are isostructural and have approximately the same unit cell dimensions. Since **1-Tb** has a $T_C$ of 423 K, **2-Dy** 393 K, and **3-Ho** the lowest $T_C$ at 370 K, this indicates a corresponding decrease in $D$ along the series. Furthermore, the exchange coupling between lanthanide-based spins and metallic electrons in an RKKY type interaction, such as in $LnMn_6Sn_6$, is proportional to the de Gennes factor, $\xi = (g_J - 1)^2 J(J + 1)$, where $g_J$ is the Landé g factor and $J$ is the total angular momentum of the $Ln^{3+}$ ion.[57] We find that the zero-field magnon frequency scales proportionally with the de Gennes factor in these materials, illustrating the strong influence of the lanthanide ion electronic configuration on the collective magnetic excitations in this family.

The last material parameter that influences the magnon frequency is the out-of-plane magnetic anisotropy of the material. As discussed in the introduction, in the $LnMn_6Sn_6$ family, the material anisotropy is controlled by the anisotropy of the $Ln^{3+}$ ion. Of the three ions studied here, the $Tb^{3+}$ ion is the most anisotropic, the $Ho^{3+}$ ion is the most isotropic and $Dy^{3+}$ is intermediate between them (Figure 4c).[48,52,53] This influences the magnetic ground state such that the magnetic ordering and spin reorientation temperatures trend with lanthanide anisotropy (i.e. **1-Tb** >> **2-Dy** > **3-Ho**). Based on our BLS results, we conclude this anisotropy also controls the magnetic excited states. Indeed, when we analytically simulate the magnon frequencies of the three materials using the Kittel equation (Figure 4b), we find the best fit to the data with an $H_{A,\,out}$ of 119,000 Oe (11.9 T) for **1-Tb**, 34,000 Oe (3.4 T) for **2-Dy**, and 24,000 Oe (2.4 T) for **3-Ho** (Table 1). Thus, we conclude that just as the anisotropy dictates the preferred orientation of the magnetization, it similarly controls the frequency of spin wave excitations.

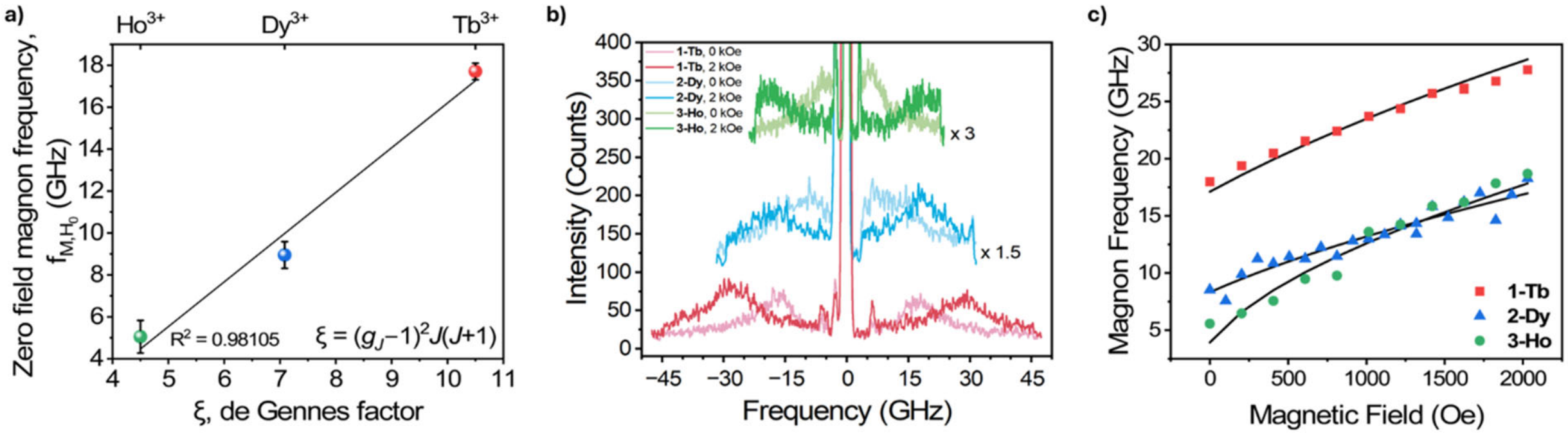

Figure 4: a) Zero field magnon frequencies plotted versus the de Gennes factor for the corresponding $Ln^{3+}$ ion. b) Offset BLS spectra of **1-Tb** (red), **2-Dy** (blue), and **3-Ho** (green) at 0 and 2 kOe. c) Experimental field dependences of the magnon frequencies of **1-Tb** (red squares), **2-Dy** (blue triangles), and **3-Ho** (green circles). Simulated data using the parameters in Table 1 are shown as black lines.

Interestingly, this trend is not reflected in the gyromagnetic ratio, $\gamma$, as **3-Ho** has the highest $\gamma$ of 2.4 x $10^6$ Hz/Oe, while **1-Tb** and **2-Dy** have values of 1.5 x $10^6$ Hz/Oe and 1.7 x $10^6$ Hz/Oe, respectively (Figure 4 and Table 1). These differences can be seen clearly in Figure 4 as the **3-Ho** dataset has a steeper slope than both **1-Tb** and **2-Dy**. In lanthanides, where spin-orbit coupling and orbital angular momentum are large, the gyromagnetic ratio can deviate strongly from the free electron value of 2.8 x $10^6$ Hz/Oe, as is observed for all the materials here.

Table 1: Simulation parameters for magnon frequency-field relationships.

| | **1-Tb** | **2-Dy** | **3-Ho** |
|---|---|---|---|
| **$H_{A,out}$ (Oe)** | 104,000 | 33,000 | 24,000 |
| **$H_{ex}$ (Oe)** | 900 | 700 | 450 |
| **$\gamma$ (Hz/Oe)** | 1.47 x $10^6$ | 1.71 x $10^6$ | 2.38 x $10^6$ |
| **$g_{eff}$** | 1.05 | 1.22 | 1.70 |

In collinear ferrimagnets, where the net magnetization stems from two antiferromagnetically coupled magnetic sublattices, the observed gyromagnetic ratio (and by extension the electronic *g* tensor since the gyromagnetic ratio and the *g*-tensor are directly related by:

$$\text{eq. (2)} \quad \gamma \;=\; g * \mu_B/h$$

where $\mu_B$ is the Bohr magneton and h is Planck's constant),[55] is determined by the ratio of the net magnetization to the net angular momentum. For the $LnMn_6Sn_6$ materials, the total effective *g*-tensor is given by the Tsuya-Wangsness formula as:

$$\text{eq. (3)} \quad g_{eff} = \frac{(M_{Mn} - M_{Ln})}{\left(\frac{M_{Mn}}{g_{Mn}} - \frac{M_{Ln}}{g_{Ln}}\right)}$$

where $M_{Mn}$ ($M_{Ln}$) is magnetization of the Mn (Ln) sublattice and $g_{Mn}$ ($g_{Ln}$) is the *g*-tensor for the Mn (Ln) sublattice.[58,59] Since the Mn sublattice is unchanged across the series **1-Tb**, **2-Dy**, and **3-Ho**, all differences in the observed $g_{eff}$ must then stem from changes in the magnetization and *g*-tensor of the lanthanide sublattice. As noted above, the magnetization of the lanthanide sublattice is largely constant between the three ions here, which means that any observed differences in $g_{eff}$ must stem primarily from the differences in the *g*-tensor of the lanthanide sublattice.

In lanthanides, the electron *g*-tensor observed in electron paramagnetic and ferromagnetic resonance spectroscopies (and thus by extension, Brillouin light scattering) is influenced by the lanthanide electronic ground state as well as the low-lying excited states induced by the crystal field splitting, the Landé *g* factor, and the total angular momentum of the ion. For example, the $Ho^{3+}$ ion has a spin-orbit coupled ground state represented by its term symbol, $^5I_8$, ($S$ = 2, $L$ = 6, $J$ = 8) that is 17-fold (2$J$ + 1) degenerate in the absence of any crystal field. In a crystal field, however, this degeneracy is lifted, and producing crystal field generated sublevels, $m_J$ = ±8, ±7, ±6, ±5, ±4, ±3, ±2, ±1, and 0. For axially symmetric systems (as is the case in $LnMn_6Sn_6$), the $\pm m_J$ doublets remain degenerate, resulting in a crystal field ground state with eight crystal field excited doublets (except in the case of $m_J$ = 0, which is a singlet).[52] Each of these sets of sublevels is affected differently by the crystal field, leading to energy separations on the order of $10^1$ – $10^2$ $cm^{-1}$ between different $\pm m_J$ states.[48,52,54] The magnitude of $m_J$ for the ground crystal field state, the presence of energetically accessible excited crystal field states, and the magnitude of $m_J$ of those excited states all influence the $g$-tensor. For axial symmetric environments, the parallel component of the $g$-tensor is given by:

$$\text{eq. (4)}\ g_{||} = 2 * g_J * M_J$$

where $M_J$ is the total angular momentum of the ion and $g_J$ is the Landé g factor,

$$\text{eq. (5)}\ g_J = 1 + \frac{J(J+1)+S(S+1)-L(L+1)}{2J(J+1)}$$

with S, L, and J representing the spin, orbital angular, and total angular momentum quantum numbers, respectively.[55] While the perpendicular component of the $g$-tensor is formally zero for strictly axial environments, a finite $g_{\perp}$ can be induced by mixing between the ground and low-lying excited states and by higher order crystal field operators that introduce off diagonal elements to the $g$-tensor. The effects of these higher order perturbations on the $g$-tensor are similarly influenced by the angular momentum of the states involved.[55] Thus while $Ho^{3+}$ has the smallest Landé $g$ factor (1.25) of the three lanthanide ions studied here, it can have the highest potential value of its electronic $g$-tensor ($g_{max}$ of 20) since it has the highest magnitude $m_J$ states. In contrast, $Tb^{3+}$ can only host a $g_{max}$ of 18.[55] This is consistent with **3-Ho** having the highest $g$-tensor and gyromagnetic ratio observed here. This has also been observed experimentally in molecular species, where in multiple series of isostructural lanthanide complexes, the identity of the lanthanide controls $g$ such that $g$ (Ho) > $g$ (Tb) > $g$ (Dy).[56]

## 3. *Conclusions*

In summary, we have investigated the magnon dynamics of $TbMn_6Sn_6$, $DyMn_6Sn_6$, and $HoMn_6Sn_6$ using Brillouin light scattering, revealing a significant influence of the lanthanide ion on the materials' magnetic excitations. This work represents the first comparative analysis of excited magnetic states in these materials and demonstrates that the magnon frequency is primarily governed by the magnetic anisotropy of the lanthanide ion, while the gyromagnetic ratio is related to the total angular momentum. These findings provide valuable insights into the relationship between the lanthanide's electronic structure and the resulting spin wave behavior in the $LnMn_6Sn_6$ family. Importantly, the tunability of magnon properties through lanthanide substitution highlights the opportunities for tailoring these materials for specific magnonic or spintronic applications. Future work in exploring the effects of alloying different lanthanides within the $LnMn_6Sn_6$ structure to achieve finer control over the magnon spectrum and investigating the temperature dependence of the magnon dynamics in greater detail, particularly in relation to the spin reorientation transitions, will further enhance this understanding and guide the design of next-generation functional magnetic materials.

## 4. *Methods*

We synthesized single crystals of $LnMn_6Sn_6$ using a Sn-based flux growth procedure modified from Clatterbuck and Gschneidner.[24] We combined high purity elements (Tb Alfa Aesar, 99.9%; Dy Alfa Aesar, 99.9%; Ho Alfa Aesar, 99.9%; Mn Alfa Aesar 99.95%; Sn Alfa Aesar 99.99%) in the ratio $(LnMn_6)_{4.5}Sn_{95.5}$ to form a ~12 g mass in a sealed Ta crucible (1.25 cm diameter, 7 cm length). Prior to sealing under partial Ar pressure with a Bühler MAM arc melter, we inserted a Ta frit to strain/separate the flux from the crystals. We then sealed the Ta assembly in an appropriately sized quartz ampoule (with quartz wool protecting the top and bottom) and placed into a box furnace inside of an $Al_2O_3$ crucible. We ramped the furnace to 1220 ºC at a rate of 50 ºC/hr, held at that temperature for ~10 hrs, and then cooled the crucible, slowly, at a rate of 1.5-2 ºC/hr until a temperature of 600 ºC was reached. At this temperature we extracted the material from the furnace, rotated it 180º, and placed in a centrifuge at high speed for 30 s to decant the Sn flux. After letting the sample cool and scoring the quartz to extract the inner sealed Ta tube, we opened the Ta tube with a standard tube cutter and carefully removed the 3 - 8 mm diameter, highly faceted crystals using dental tools. The details of a similar synthesis procedure can be found in a previous publication by Susner et al.[60] We then determined the composition of the material through EDS spectroscopy using an Hitachi TM 4000 Plus mated with Oxford Aztec One Xplore30 Compact EDS system with Live mapping. We checked for the correct structure using a Malvern PanAlytical Empyrian X-ray Diffractometer. Magnetometry data were collected on oriented single crystals of **1-Tb**, **2-Dy**, and **3-Ho** mounted on quartz paddles with magnetic field applied parallel to the crystallographic *ab* plane using the VSM option on a Quantum Design Dynacool 9 T Physical Property Measurement System (PPMS).

A micro-Brillouin scattering system with a (3 + 3) pass Sandercock-type tandem Fabry−Perot interferometer was used to measure the Brillouin spectra of **1-Tb**, **2-Dy**, and **3-Ho** in a backward scattering geometry with a free spectral range of 45, 30, and 25 GHz, respectively. Samples are bulk crystals ranging from 0.5 − 5 mm in size. A diode pumped solid state laser at the wavelength of 532 nm was used as an excitation light source. The laser spot size of the BLS microscope is approximately 3 μm. The power of the incident laser is approximately 8 mW at the sample surface, and spectra were collected for 1 hour each. Magnetic fields were applied with a GMW Model 5403 electromagnet with a Kepco BOP 20-20M power supply. For variable temperature measurements, the sample was placed inside a liquid nitrogen cooled cryostat cell (Mikroptik) and the temperature was changed from −30 °C to 100 °C.

## *Supplemental Material*

Vibrating sample magnetometry results are available in the electronic supplemental information. Complete Brillouin light scattering spectra for 1-Tb, 2-Dy, and 3-Ho and their individual Lorentzian fits are also available in the supplementary material.

## *Acknowledgements*

K.A.C, M.A.S., and M.J.N. gratefully acknowledge support from the Air Force Research Laboratory's Materials and Manufacturing Directorate and Air Force Office of Scientific Research, including Grant Nos. LRIR 23RXCOR001, LRIR 23RXCOR003, and LRIR 26RXCOR010. M.J. gratefully acknowledges the Air Force Research Laboratory, Materials and Manufacturing Directorate (AFRL/RXMS) for support via Contract No. FA8650-21-C5711.

## *Conflict of Interest*

The authors have no conflicts of interest to declare.

### *Data Availability Statement*

The data that support the findings of this study are available in the electronic supporting information and from the corresponding author upon reasonable request.

## B. References

# Supporting Information

**For:**

**Lanthanide Ion Electronic Structure Controls Magnetic Excitations in Topological Quantum Ferrimagnets $LnMn_6Sn_6$ (Ln = Tb, Dy, Ho)**

Kelsey A. Collins,[1,2*] Jacob Pfund[1,3], Michael R. Page,[1] Menka Jain,[3] Michael A. Susner,[1] and Michael J. Newburger[1]

[1]Materials and Manufacturing Directorate, Air Force Research Laboratory, Wright-Patterson Air Force Base, OH 45433, USA

[2]Core4ce, Dayton, OH 45433, USA

[3]Department of Physics, University of Connecticut, Storrs, CT 06269, USA

*corresponding author: kelsey.collins.1.ctr@afrl.af.mil

## Vibrating sample magnetometry

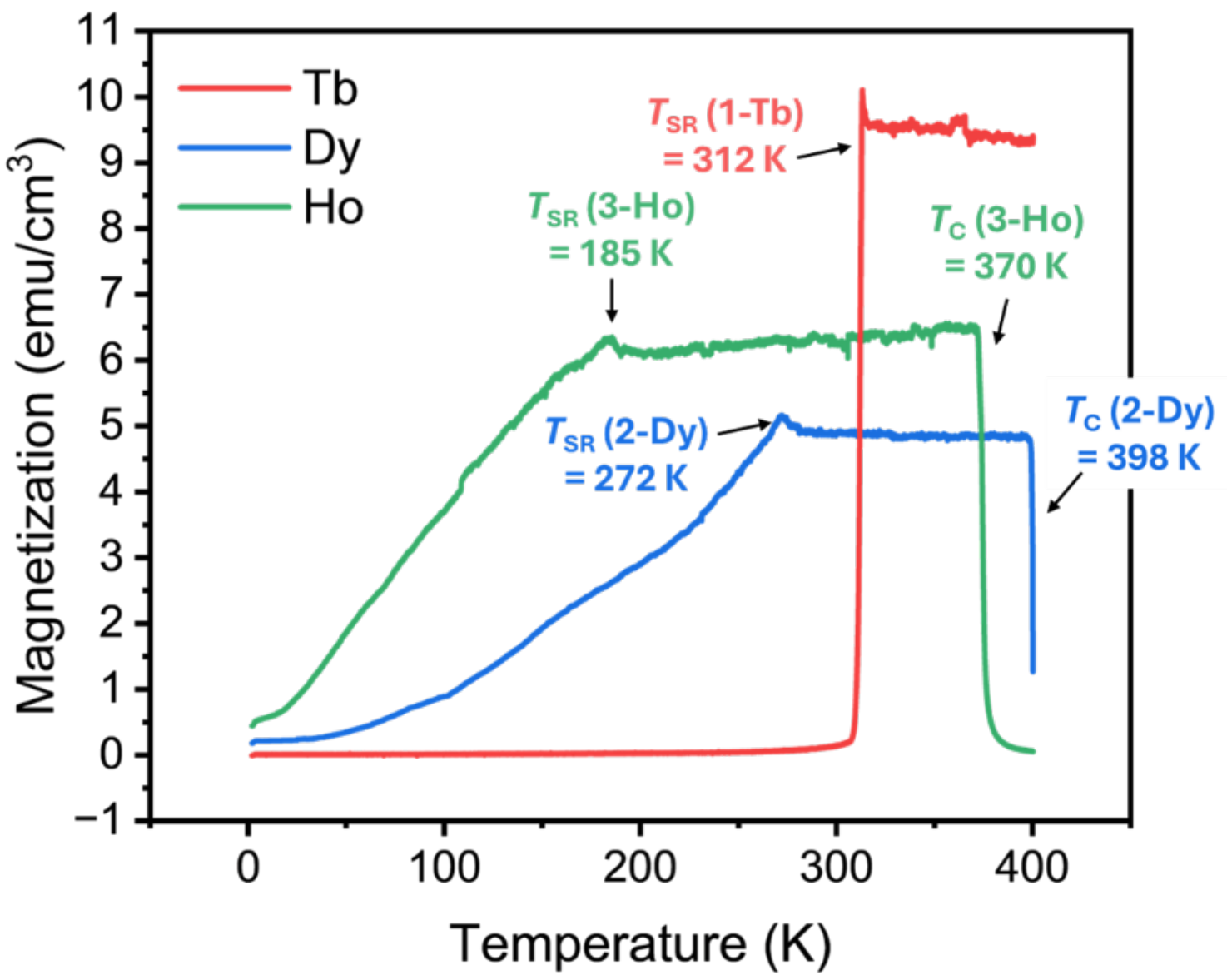

**Figure S-1**: Variable temperature magnetization from 400 – 1.8 K with H = 10 Oe applied parallel to crystallographic *ab* plane (normal to crystallographic *c* axis) for **1-Tb**, **2-Dy**, and **3-Ho** crystals, showing the spin-reorientation transition ($T_{SR}$) and ordering temperatures ($T_C$). $T_C$ for **1-Tb** is not observed due to it being above 400 K.

**Alt text**: A line graph tracking variable-temperature magnetization under a small external field. The data plots magnetization against temperature for terbium, dysprosium, and holmium compounds, illustrating their distinct spin-reorientation and magnetic ordering transition temperatures.

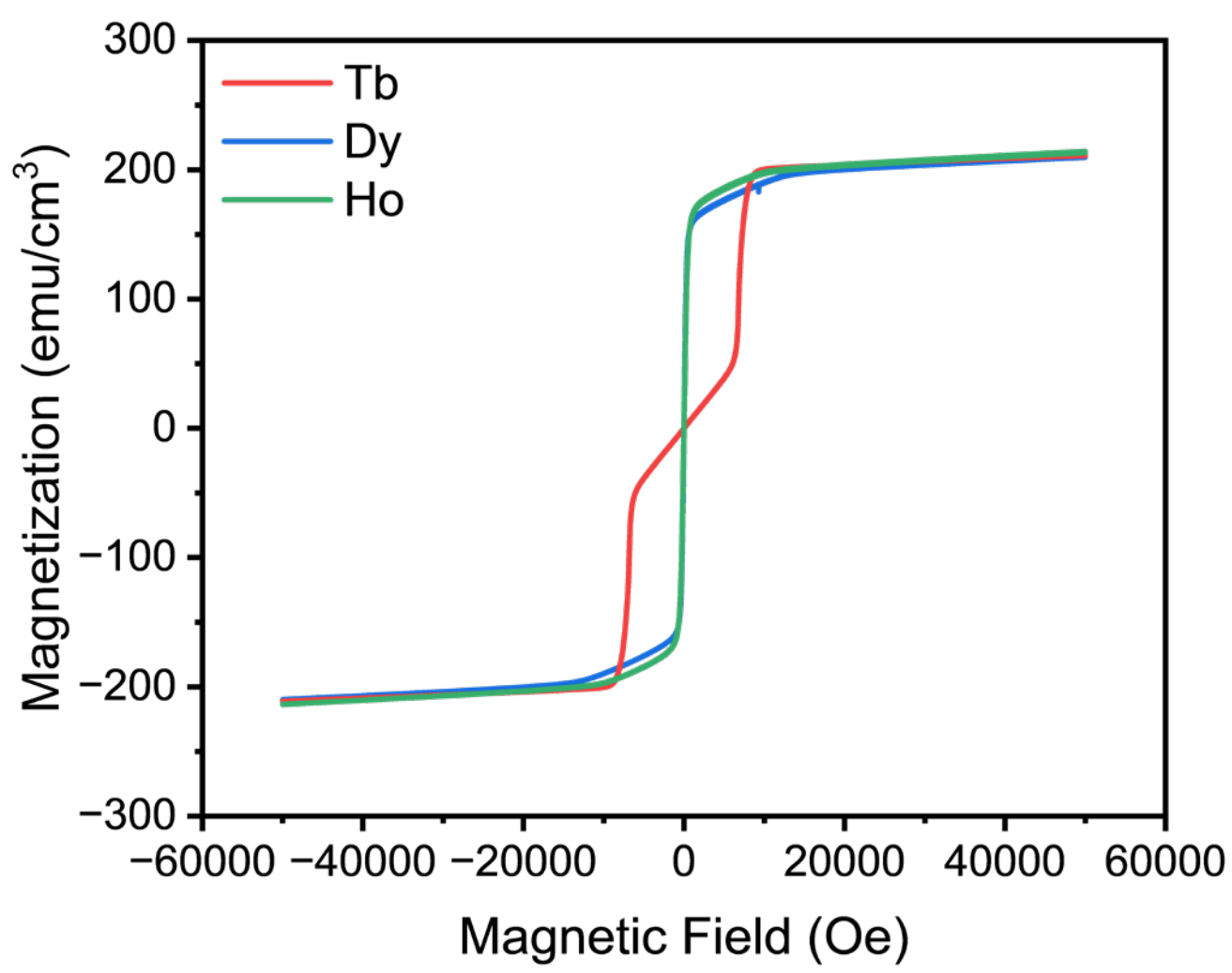

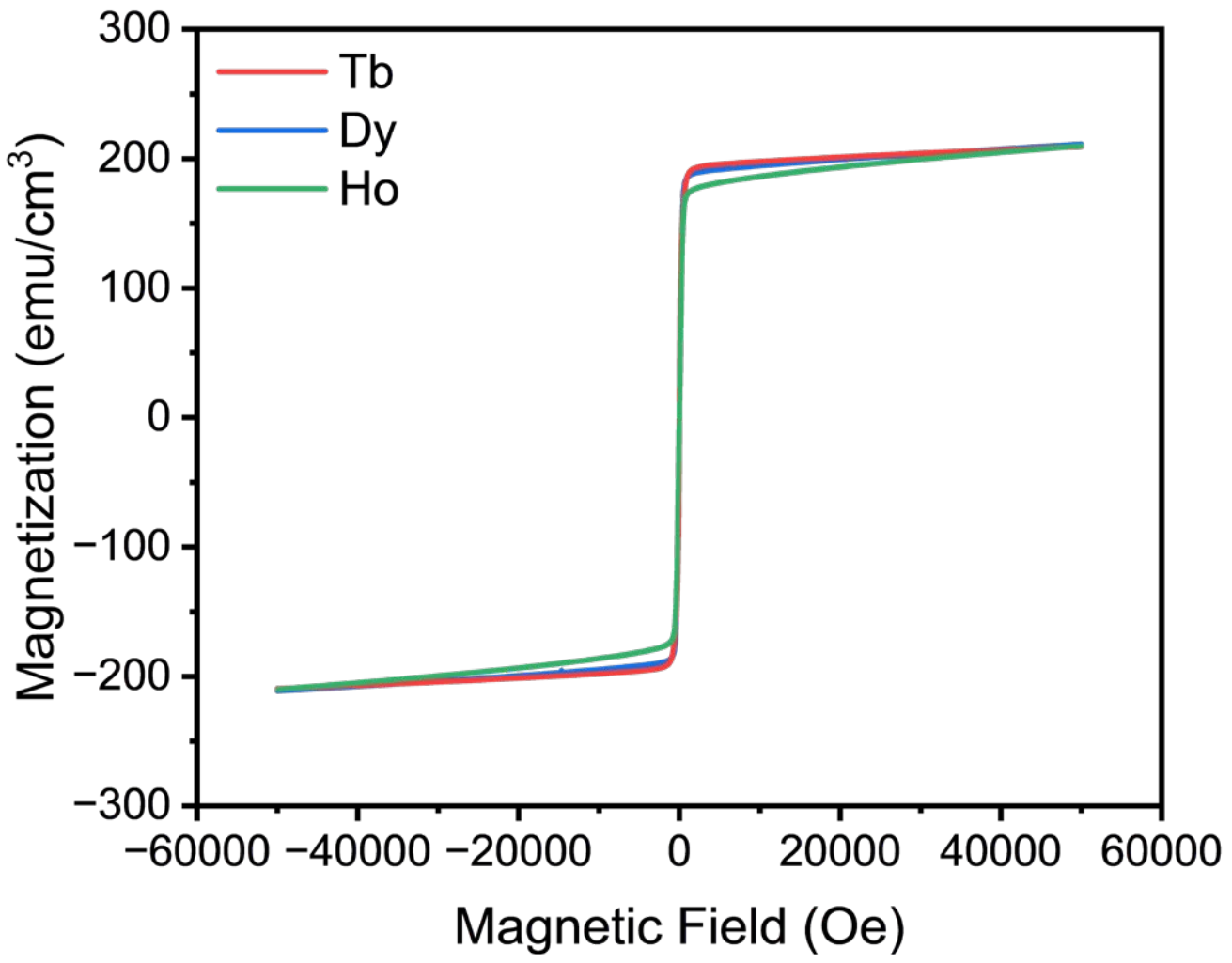

**Figure S-2**: (Top) Variable field magnetization curves measured with H || *ab* plane and at T < $T_{SR}$, i.e. below the spin-reorientation temperature of **1-Tb** (T = 290 K), **2-Dy** (T = 250 K), and **3-Ho** (T = 170 K). (Bottom) Variable field magnetization curves measured with H || *ab* plane and at T > $T_{SR}$ i.e. above the spin-reorientation temperature of **1-Tb**, **2-Dy**, and **3-Ho** (T = 320 K), showing the different saturation field dependent behaviors between T < $T_{SR}$ and T > $T_{SR}$.

**Alt text**: A two-panel compilation of variable-field magnetization curves. The top graph displays magnetic hysteresis loops measured below the spin-reorientation temperatures, while the bottom graph displays loops measured above the spin-reorientation temperatures to compare saturation field dependencies.

## Room temperature BLS spectra fits of 1-Tb

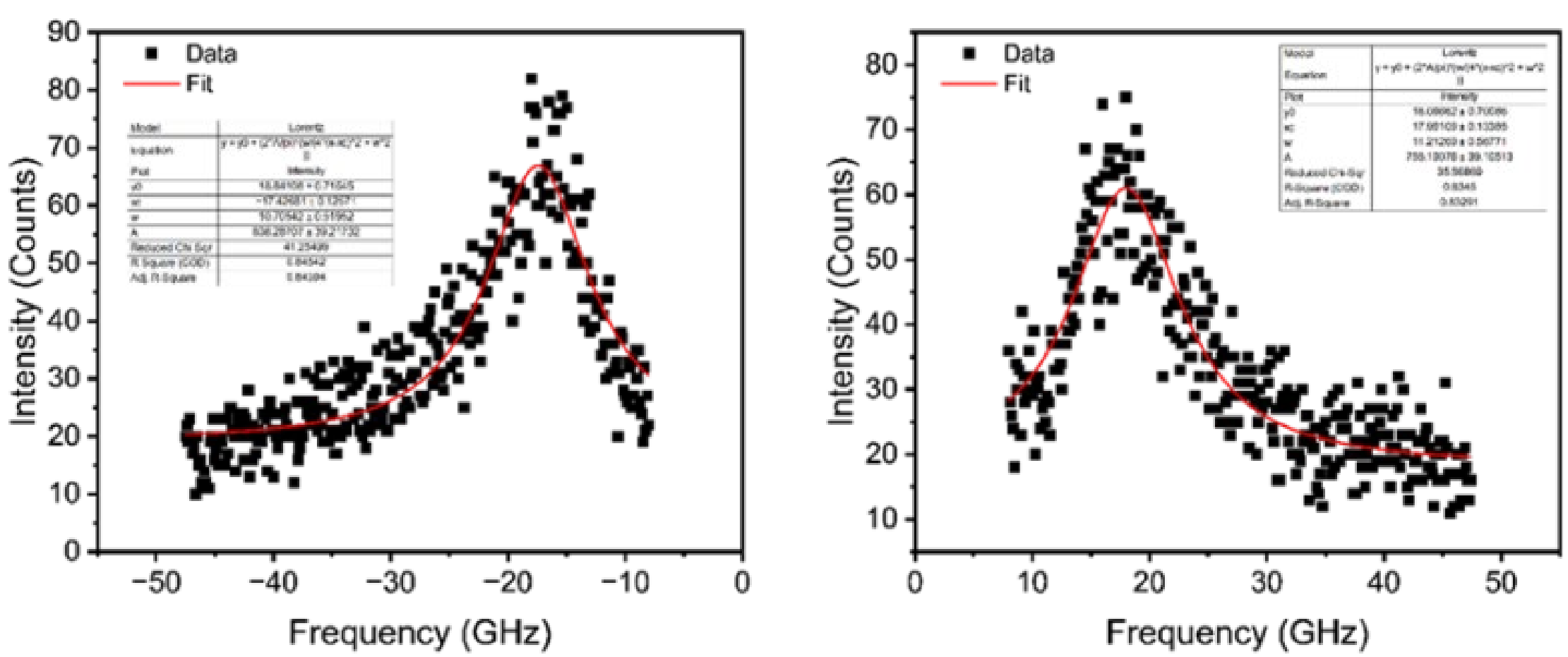

Figure S-3: Fit of $TbMn_6Sn_6$ Stokes (left) and Anti-Stokes (right) magnon at 0 mT.

Alt text: Each figure displays raw Brillouin light scattering data points fitted with a solid red Lorentzian curve, plotting scattering intensity against frequency shift to resolve the respective Stokes and anti-Stokes magnon peaks at incremental magnetic field values.

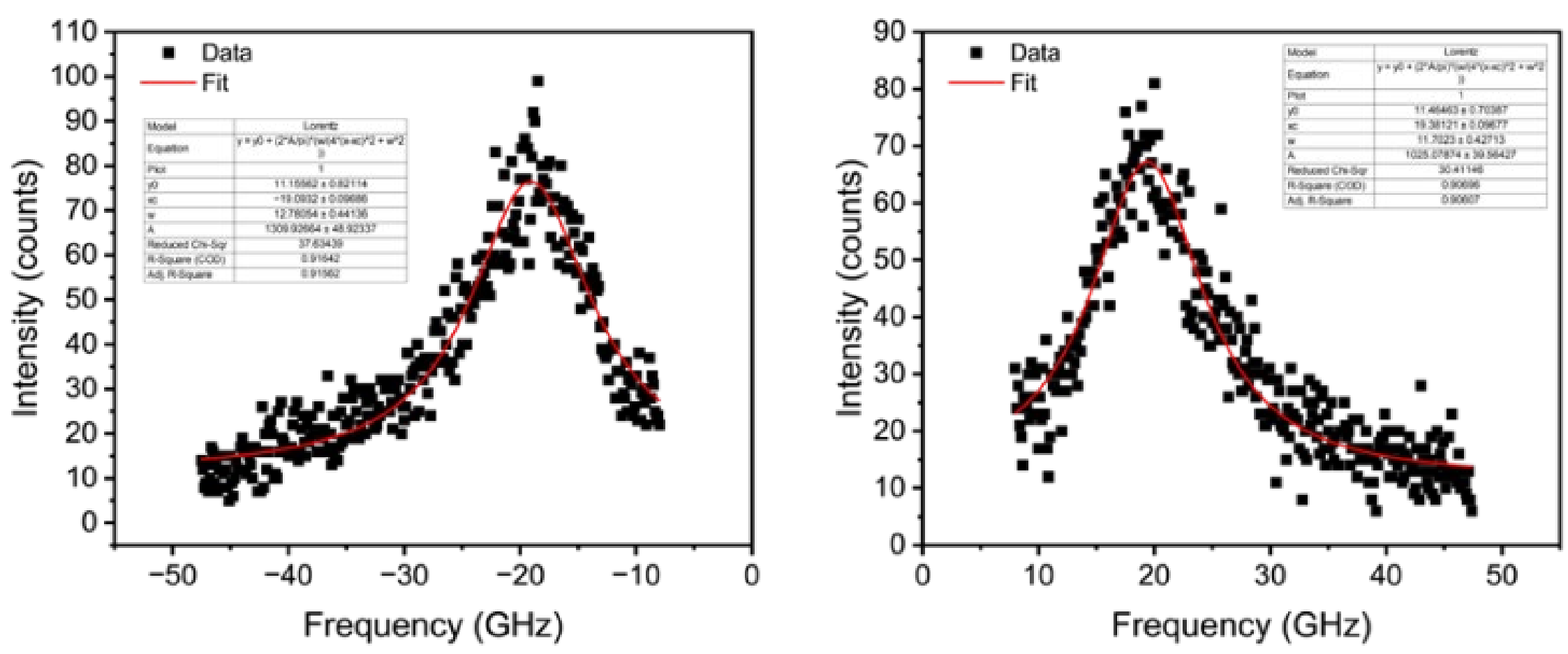

Figure S-4: Fit of $TbMn_6Sn_6$ Stokes (left) and Anti-Stokes (right) magnon at 20 mT.

Alt text: Each figure displays raw Brillouin light scattering data points fitted with a solid red Lorentzian curve, plotting scattering intensity against frequency shift to resolve the respective Stokes and anti-Stokes magnon peaks at incremental magnetic field values.

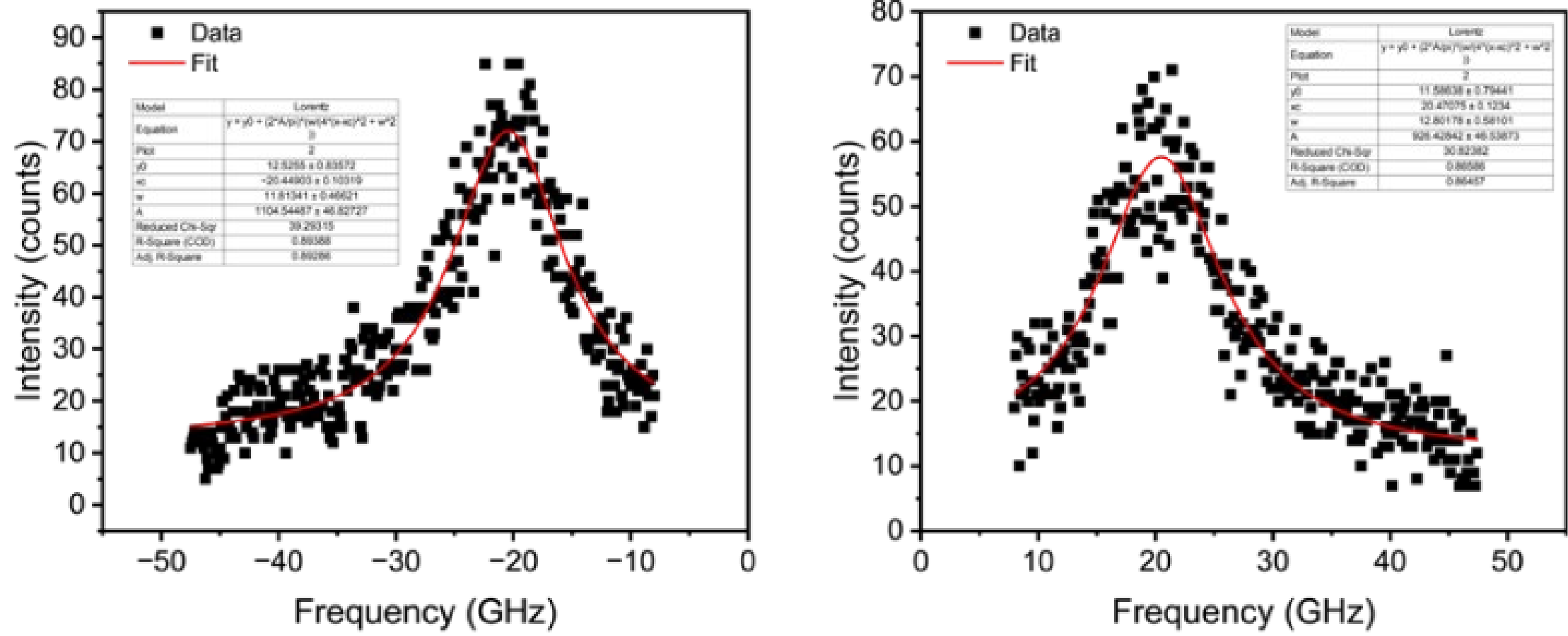

Figure S-5: Fit of $TbMn_6Sn_6$ Stokes (left) and Anti-Stokes (right) magnon at 40 mT.

Alt text: Each figure displays raw Brillouin light scattering data points fitted with a solid red Lorentzian curve, plotting scattering intensity against frequency shift to resolve the respective Stokes and anti-Stokes magnon peaks at incremental magnetic field values.

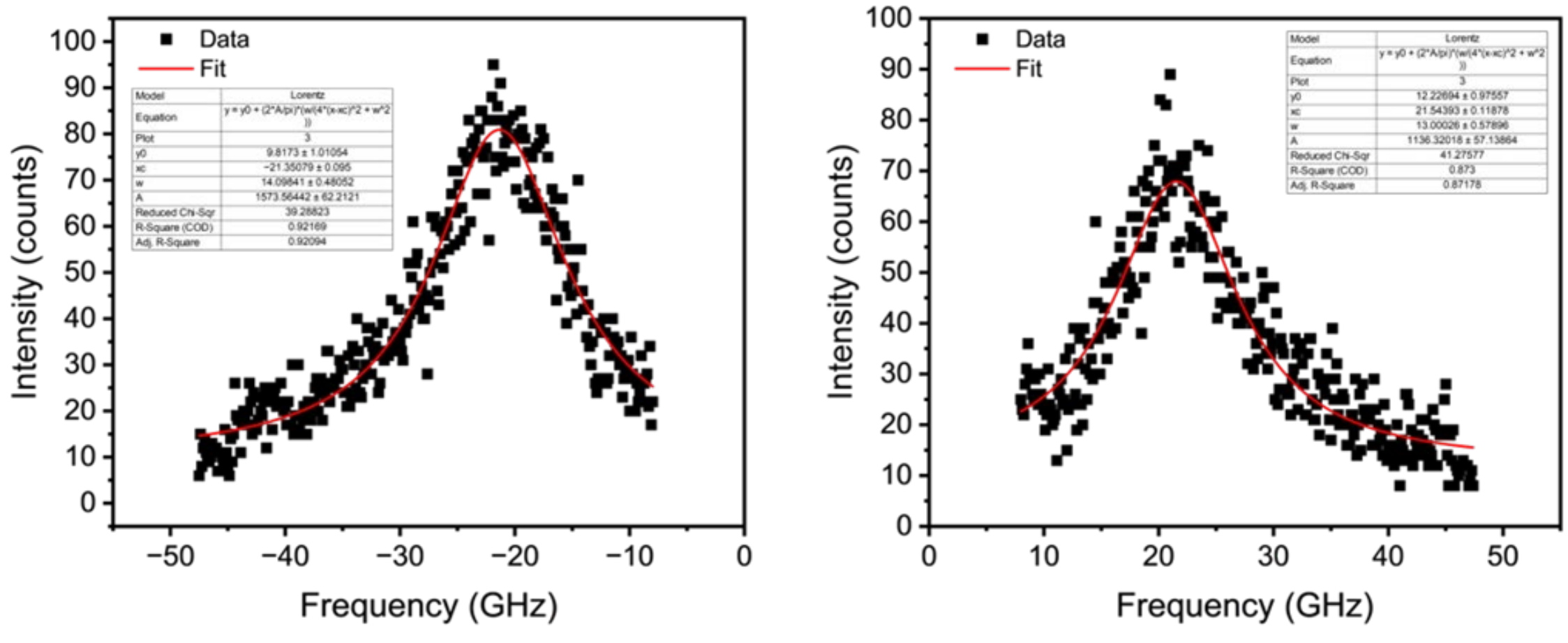

Figure S-6: Fit of $TbMn_6Sn_6$ Stokes (left) and Anti-Stokes (right) magnon at 60 mT.

Alt text: Each figure displays raw Brillouin light scattering data points fitted with a solid red Lorentzian curve, plotting scattering intensity against frequency shift to resolve the respective Stokes and anti-Stokes magnon peaks at incremental magnetic field values.

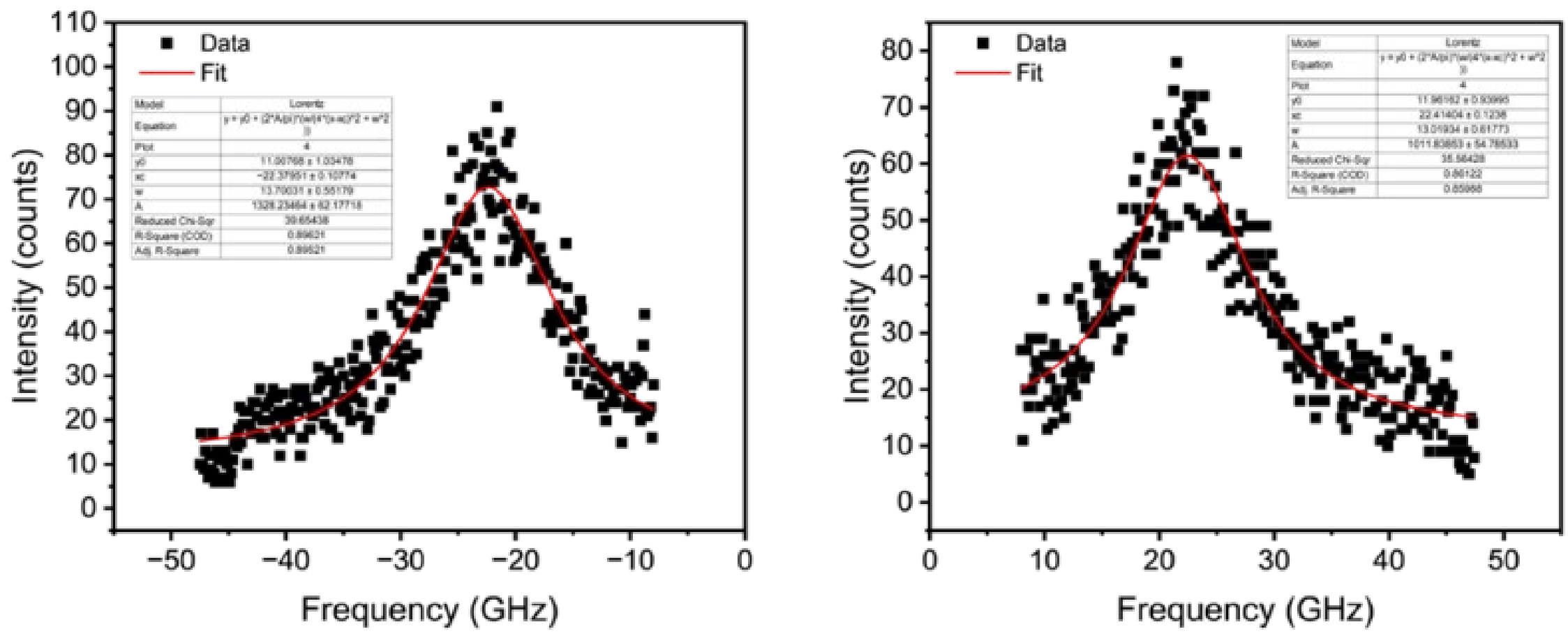

Figure S-7: Fit of $TbMn_6Sn_6$ Stokes (left) and Anti-Stokes (right) magnon at 80 mT.

Alt text: Each figure displays raw Brillouin light scattering data points fitted with a solid red Lorentzian curve, plotting scattering intensity against frequency shift to resolve the respective Stokes and anti-Stokes magnon peaks at incremental magnetic field values.

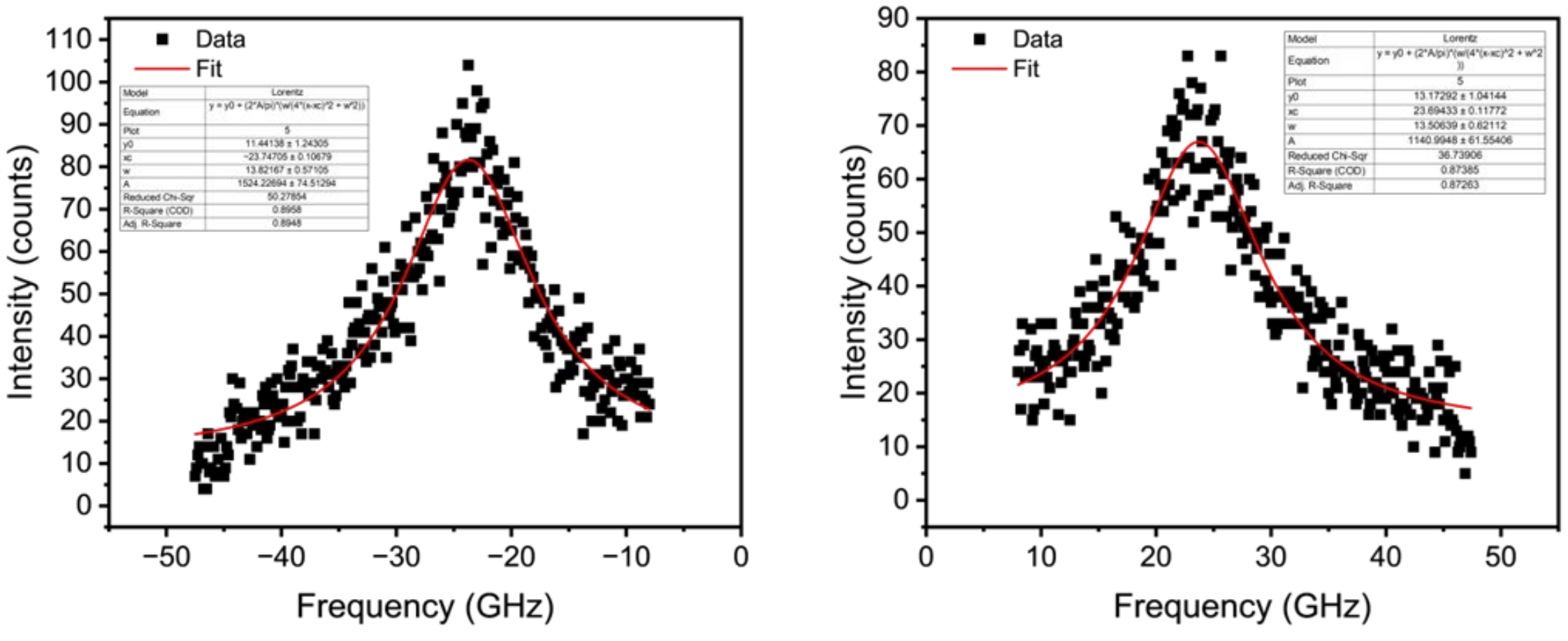

Figure S-8: Fit of $TbMn_6Sn_6$ Stokes (left) and Anti-Stokes (right) magnon at 100 mT.

Alt text: Each figure displays raw Brillouin light scattering data points fitted with a solid red Lorentzian curve, plotting scattering intensity against frequency shift to resolve the respective Stokes and anti-Stokes magnon peaks at incremental magnetic field values.

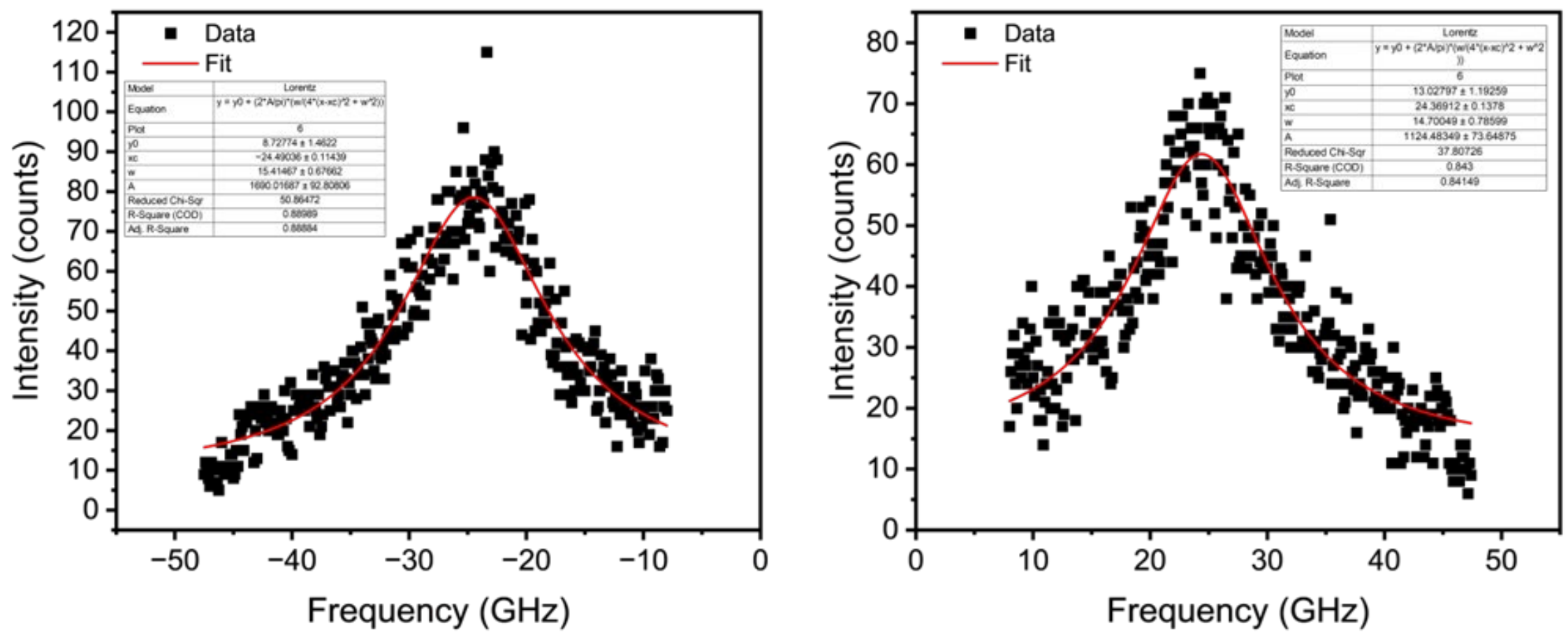

Figure S-9: Fit of $TbMn_6Sn_6$ Stokes (left) and Anti-Stokes (right) magnon at 120 mT.

Alt text: Each figure displays raw Brillouin light scattering data points fitted with a solid red Lorentzian curve, plotting scattering intensity against frequency shift to resolve the respective Stokes and anti-Stokes magnon peaks at incremental magnetic field values.

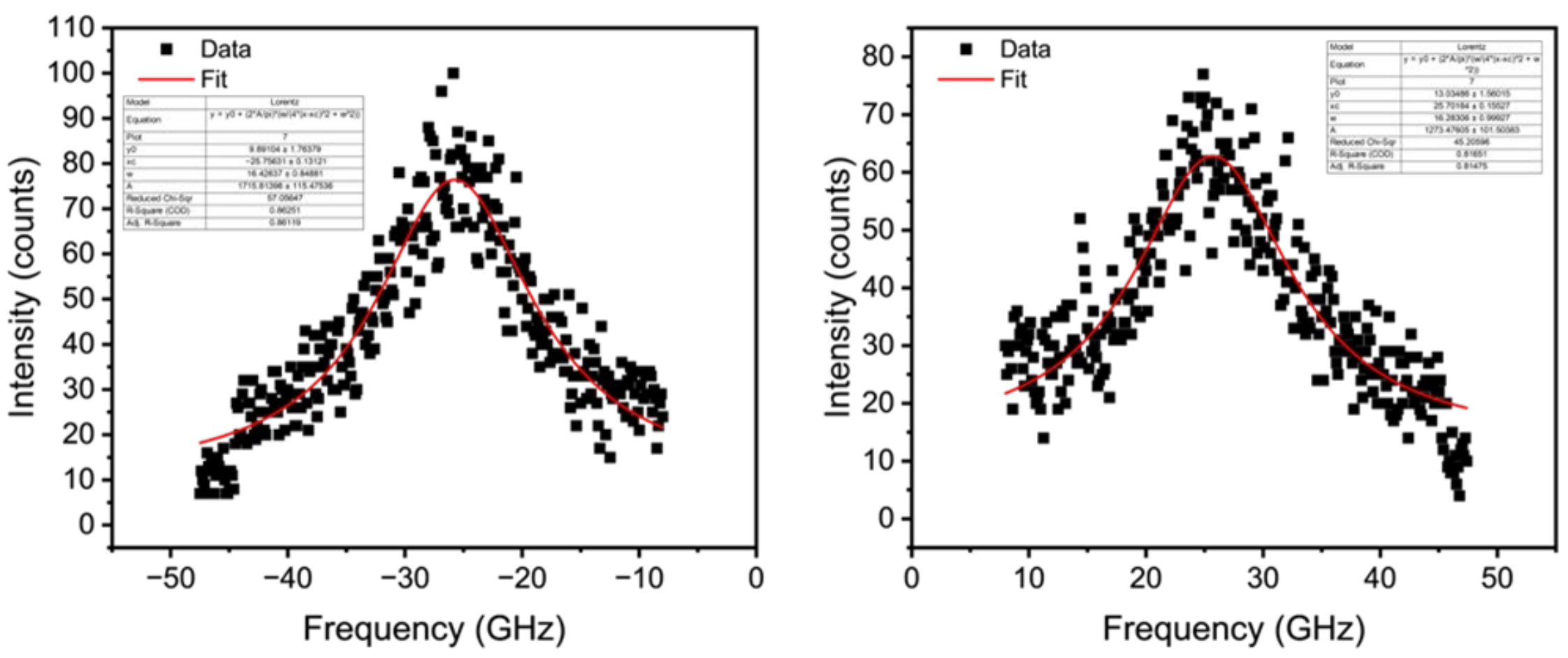

Figure S-10: Fit of $TbMn_6Sn_6$ Stokes (left) and Anti-Stokes (right) magnon at 140 mT.

Alt text: Each figure displays raw Brillouin light scattering data points fitted with a solid red Lorentzian curve, plotting scattering intensity against frequency shift to resolve the respective Stokes and anti-Stokes magnon peaks at incremental magnetic field values.

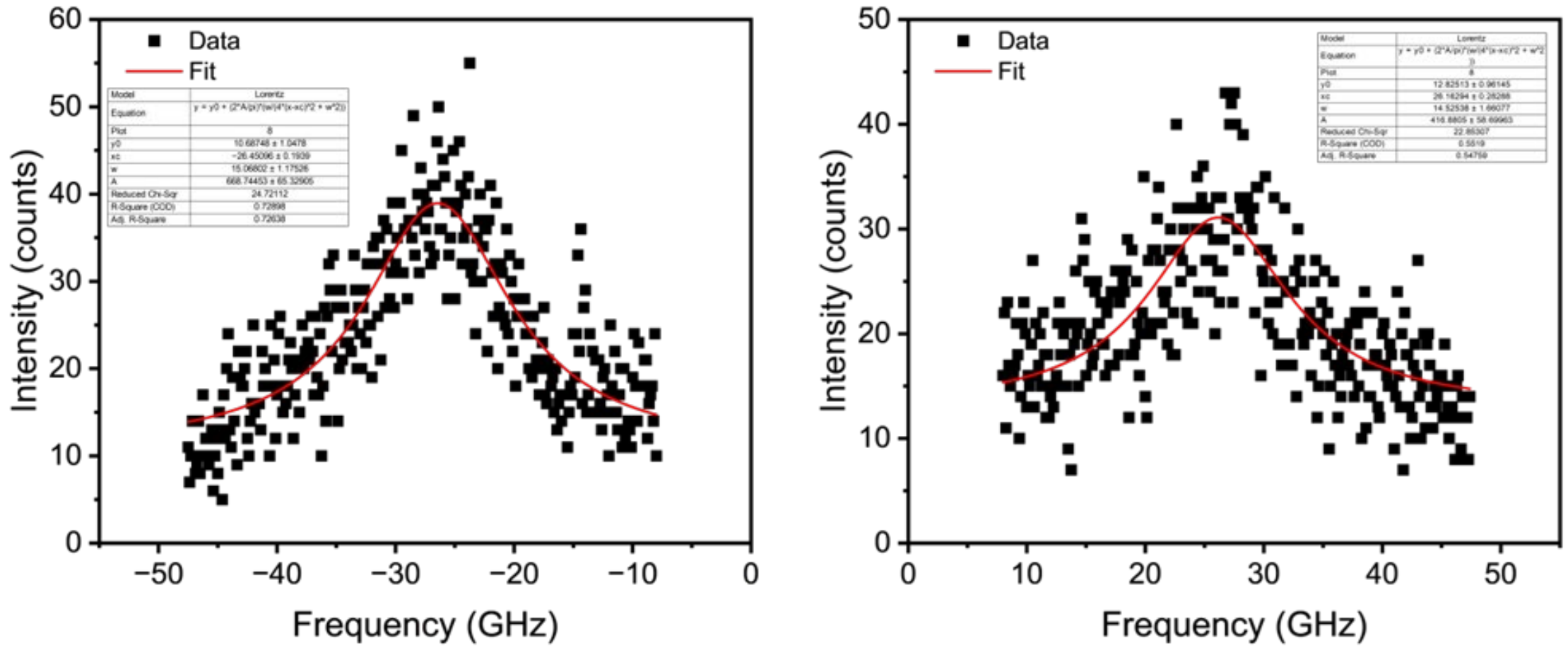

Figure S-11: Fit of $TbMn_6Sn_6$ Stokes (left) and Anti-Stokes (right) magnon at 160 mT.

Alt text: Each figure displays raw Brillouin light scattering data points fitted with a solid red Lorentzian curve, plotting scattering intensity against frequency shift to resolve the respective Stokes and anti-Stokes magnon peaks at incremental magnetic field values.

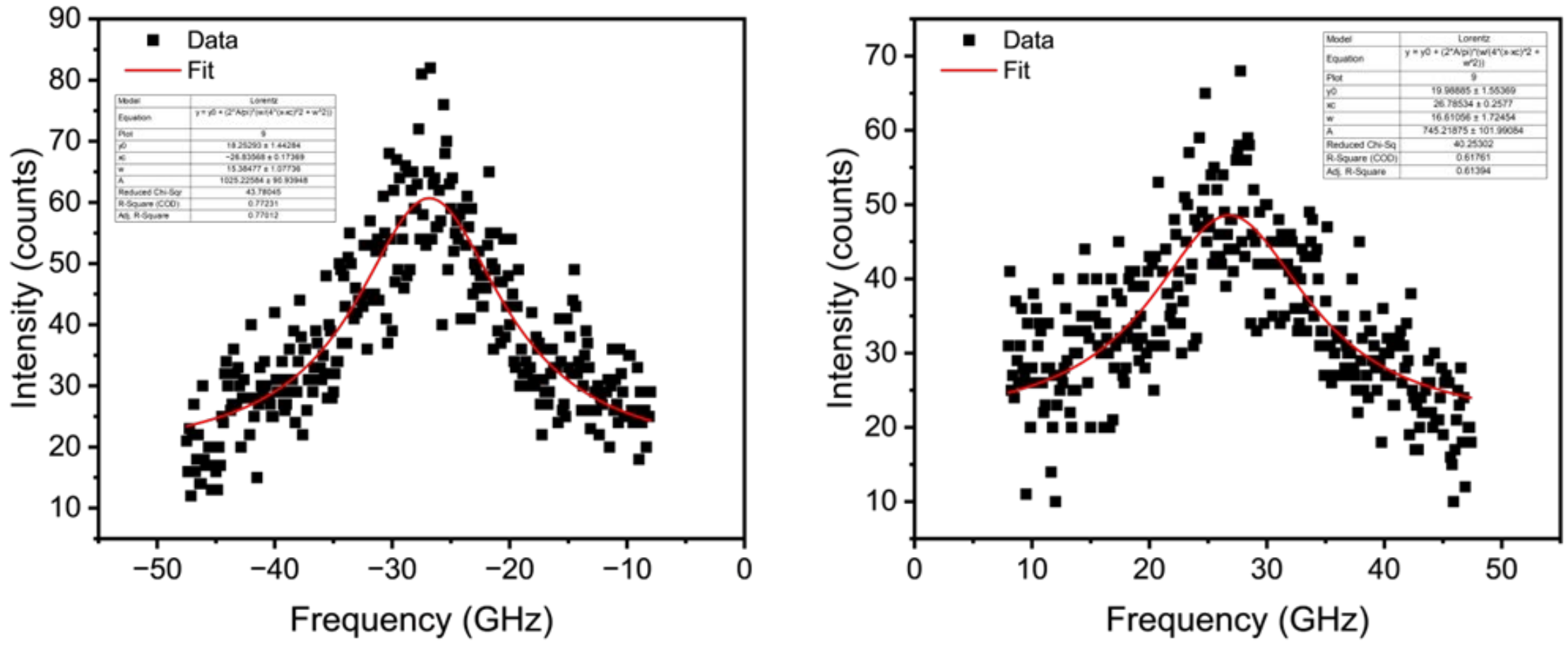

Figure S-12: Fit of $TbMn_6Sn_6$ Stokes (left) and Anti-Stokes (right) magnon at 180 mT.

Alt text: Each figure displays raw Brillouin light scattering data points fitted with a solid red Lorentzian curve, plotting scattering intensity against frequency shift to resolve the respective Stokes and anti-Stokes magnon peaks at incremental magnetic field values.

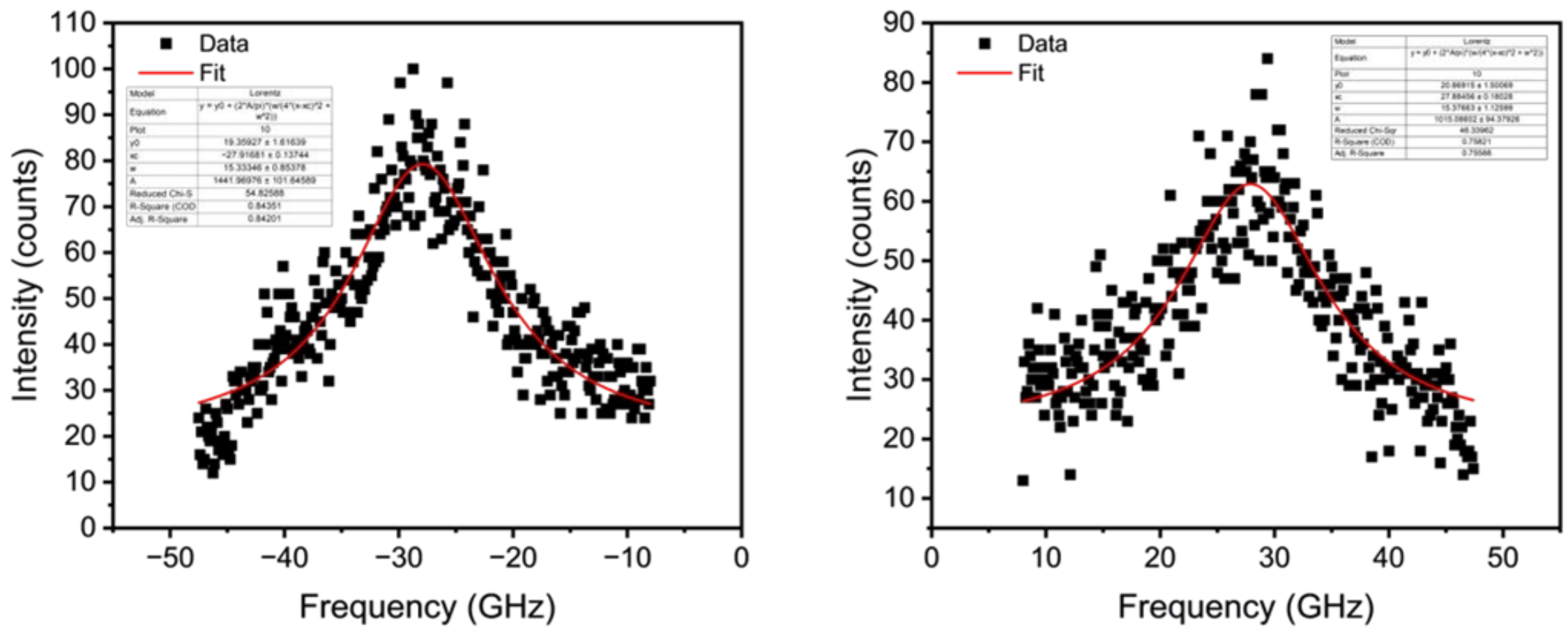

**Figure S-13**: Fit of $TbMn_6Sn_6$ Stokes (left) and Anti-Stokes (right) magnon at 200 mT.

**Alt text**: Each figure displays raw Brillouin light scattering data points fitted with a solid red Lorentzian curve, plotting scattering intensity against frequency shift to resolve the respective Stokes and anti-Stokes magnon peaks at incremental magnetic field values.

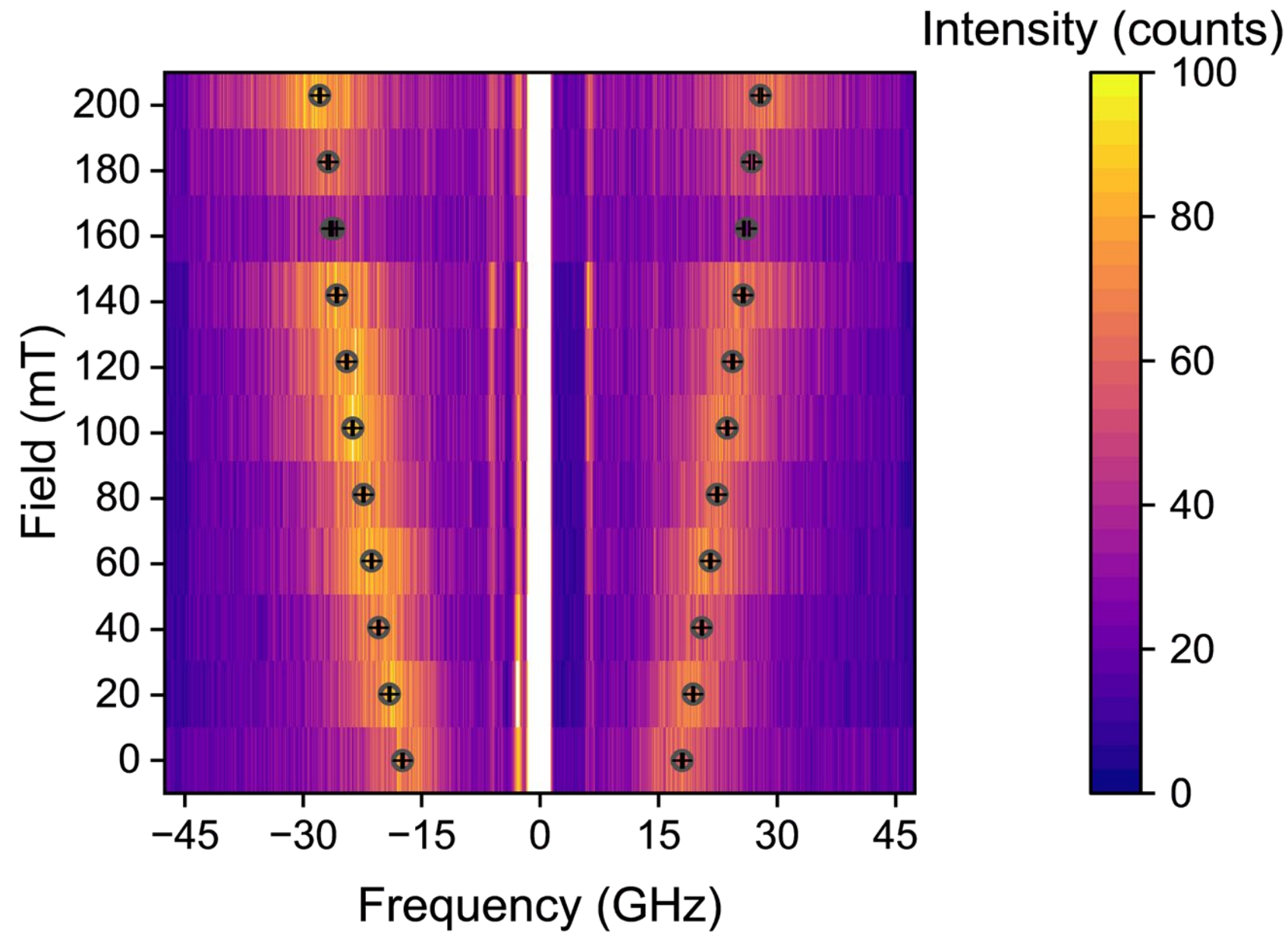

**Figure S-14**: 1-Tb heatmap with fitted magnon frequencies.

**Alt text:** A 2D intensity heatmap. The plot describes the field-swept Brillouin light scattering spectra for the terbium compound, tracking intensity variations against applied magnetic field and frequency shift, with fitted magnon positions overlaid as data points.

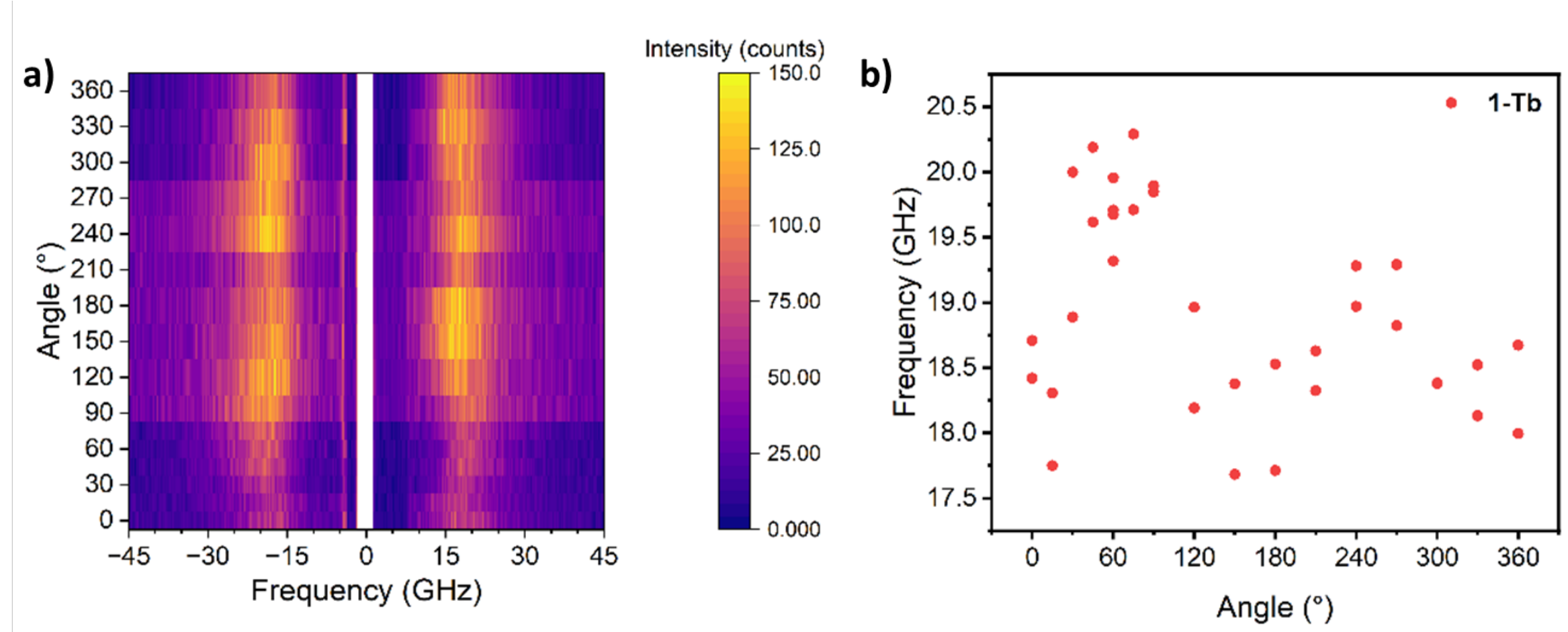

**Figure S-15**: a) In-plane angular dependence of 1-Tb magnon spectra with b) fitted magnon frequencies as a function of in-plane rotation angle.

**Alt text**: A two-panel angular dependence analysis. Panel (a) is a 2D intensity heatmap plotting light scattering intensity against frequency shift and in-plane rotation angle. Panel (b) is a scatter plot mapping the extracted magnon frequencies as a function of the in-plane rotation angle.

# Room temperature BLS spectra fits of 2-Dy

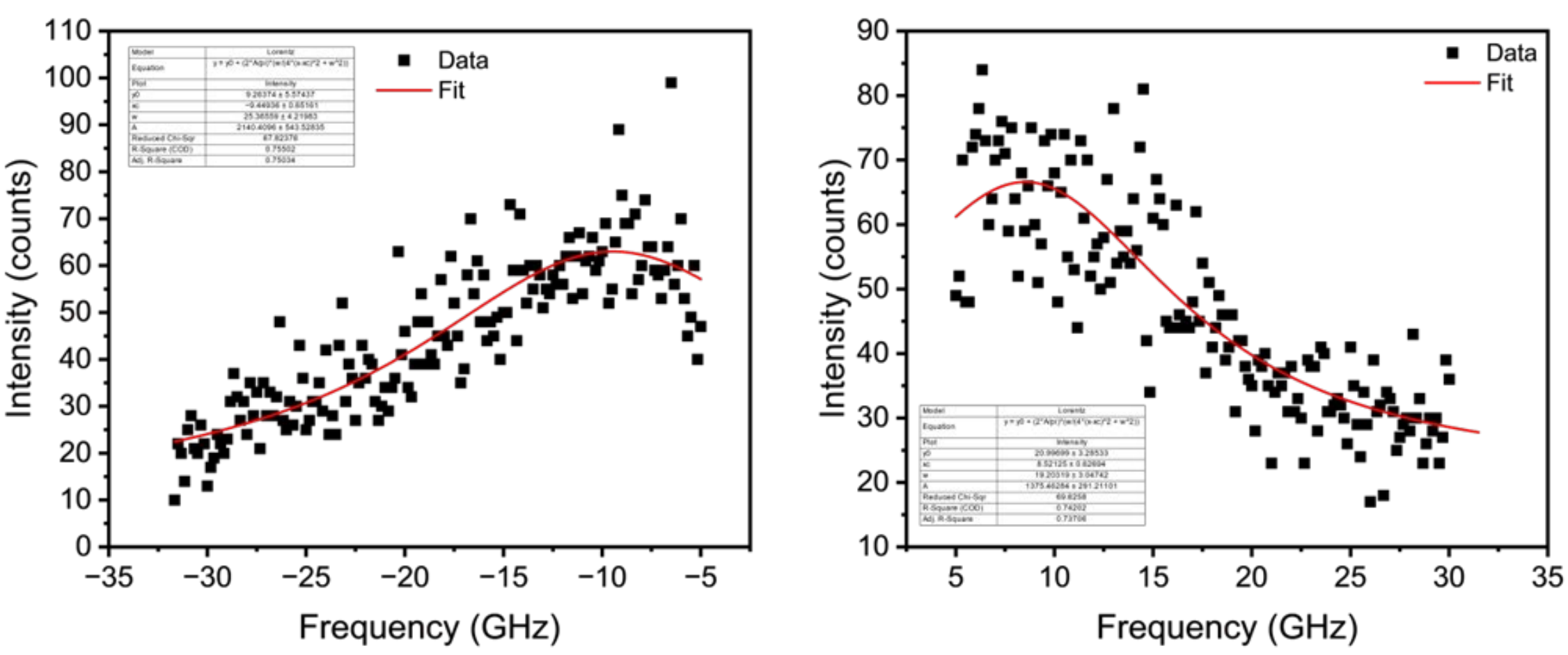

**Figure S-16**: Fit of *Dy*$Mn_6Sn_6$ Stokes (left) and Anti-Stokes (right) magnon at 0 mT.

**Alt text**: Each figure displays raw Brillouin light scattering spectra fitted with a red Lorentzian line shape, plotting intensity against frequency shift to capture the broad Stokes and anti-Stokes magnon features of the dysprosium compound across multiple field values.

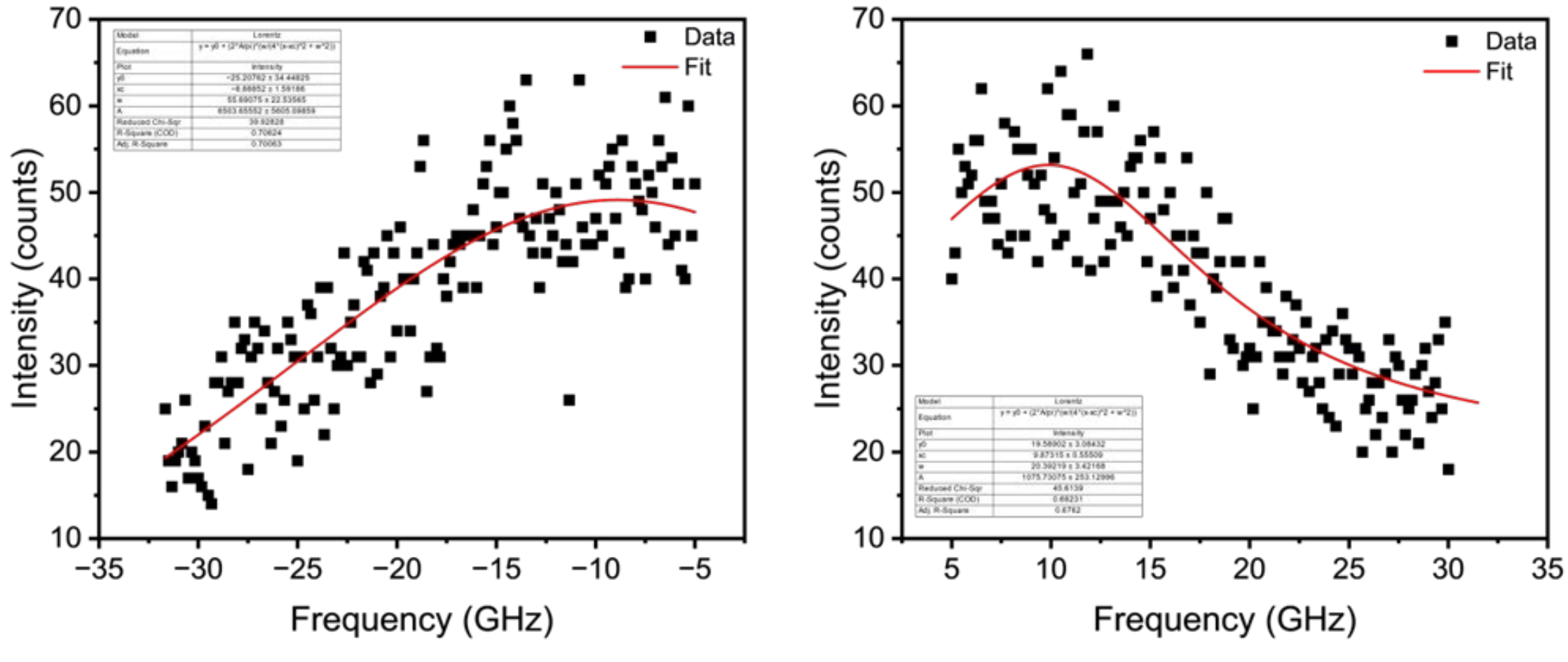

**Figure S-17**: Fit of *Dy*$Mn_6Sn_6$ Stokes (left) and Anti-Stokes (right) magnon at 20 mT.

**Alt text**: Each figure displays raw Brillouin light scattering spectra fitted with a red Lorentzian line shape, plotting intensity against frequency shift to capture the broad Stokes and anti-Stokes magnon features of the dysprosium compound across multiple field values.

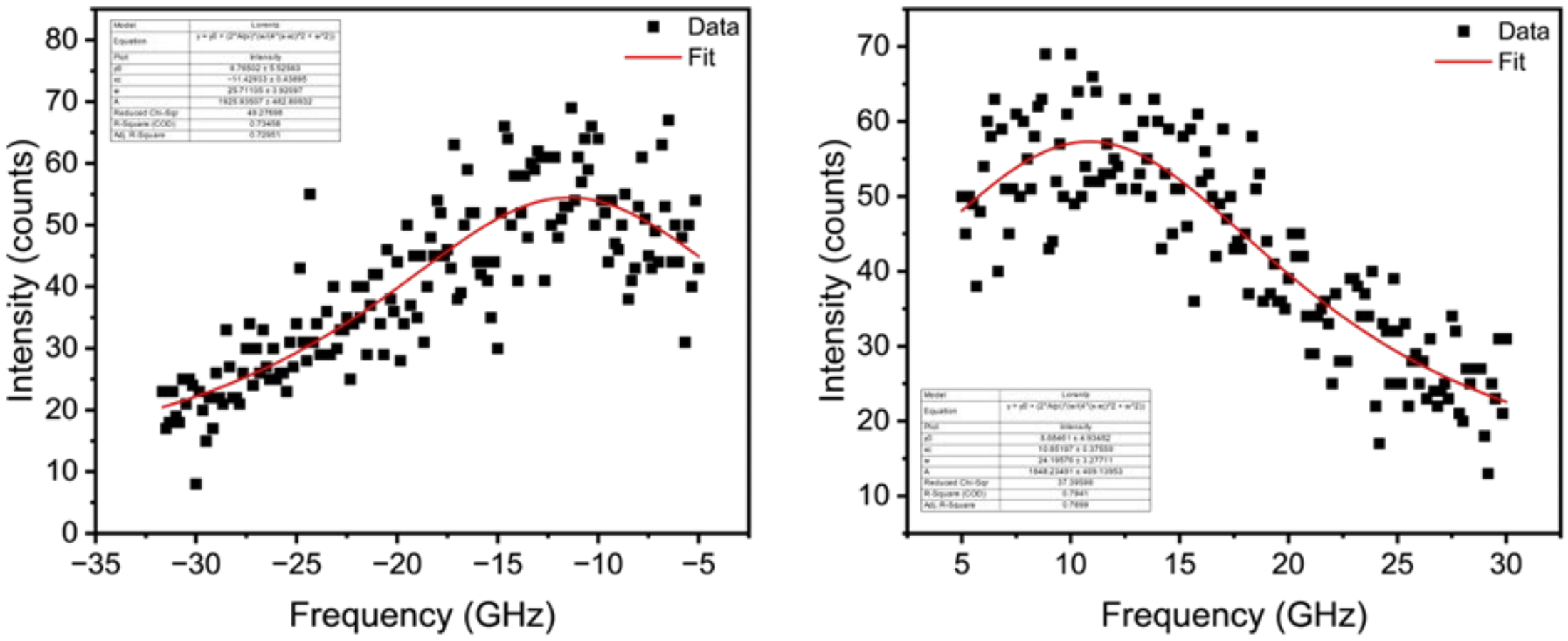

**Figure S-18**: Fit of *Dy*$Mn_6Sn_6$ Stokes (left) and Anti-Stokes (right) magnon at 40 mT.

**Alt text**: Each figure displays raw Brillouin light scattering spectra fitted with a red Lorentzian line shape, plotting intensity against frequency shift to capture the broad Stokes and anti-Stokes magnon features of the dysprosium compound across multiple field values.

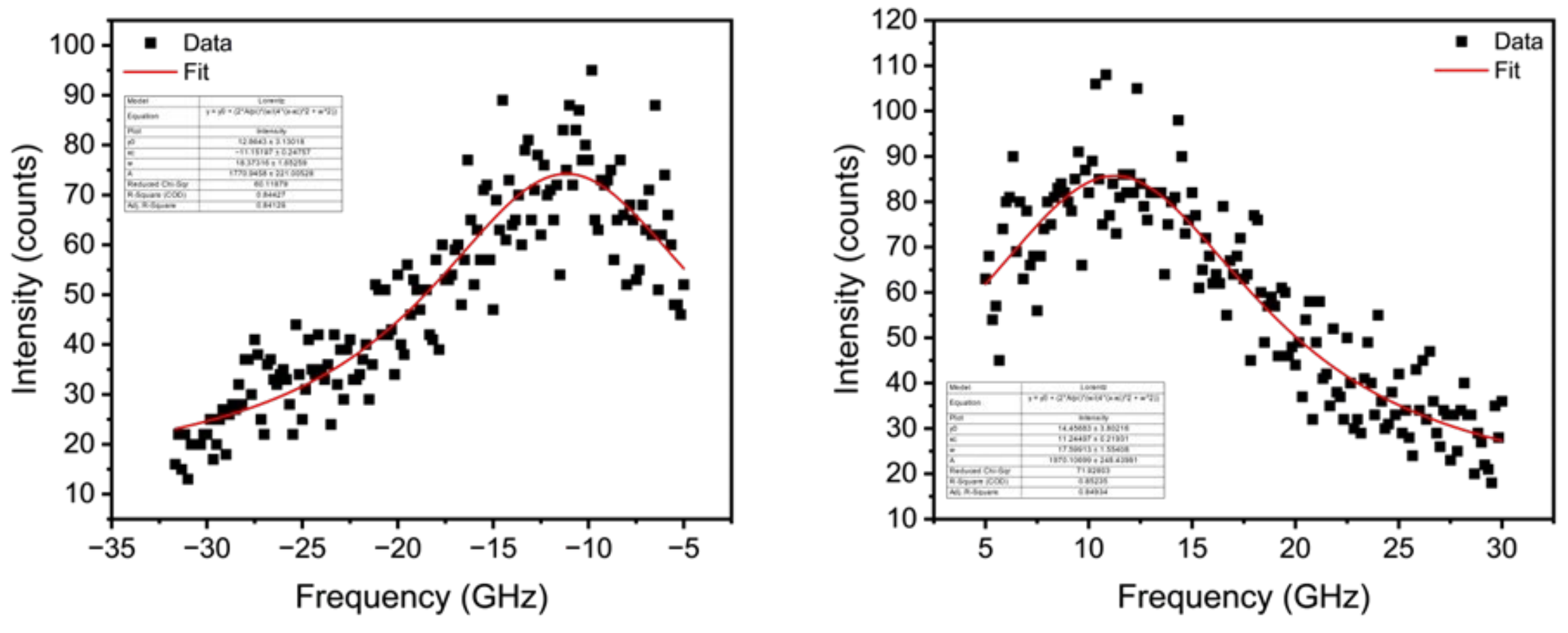

**Figure S-19**: Fit of *Dy*$Mn_6Sn_6$ Stokes (left) and Anti-Stokes (right) magnon at 60 mT.

**Alt text**: Each figure displays raw Brillouin light scattering spectra fitted with a red Lorentzian line shape, plotting intensity against frequency shift to capture the broad Stokes and anti-Stokes magnon features of the dysprosium compound across multiple field values.

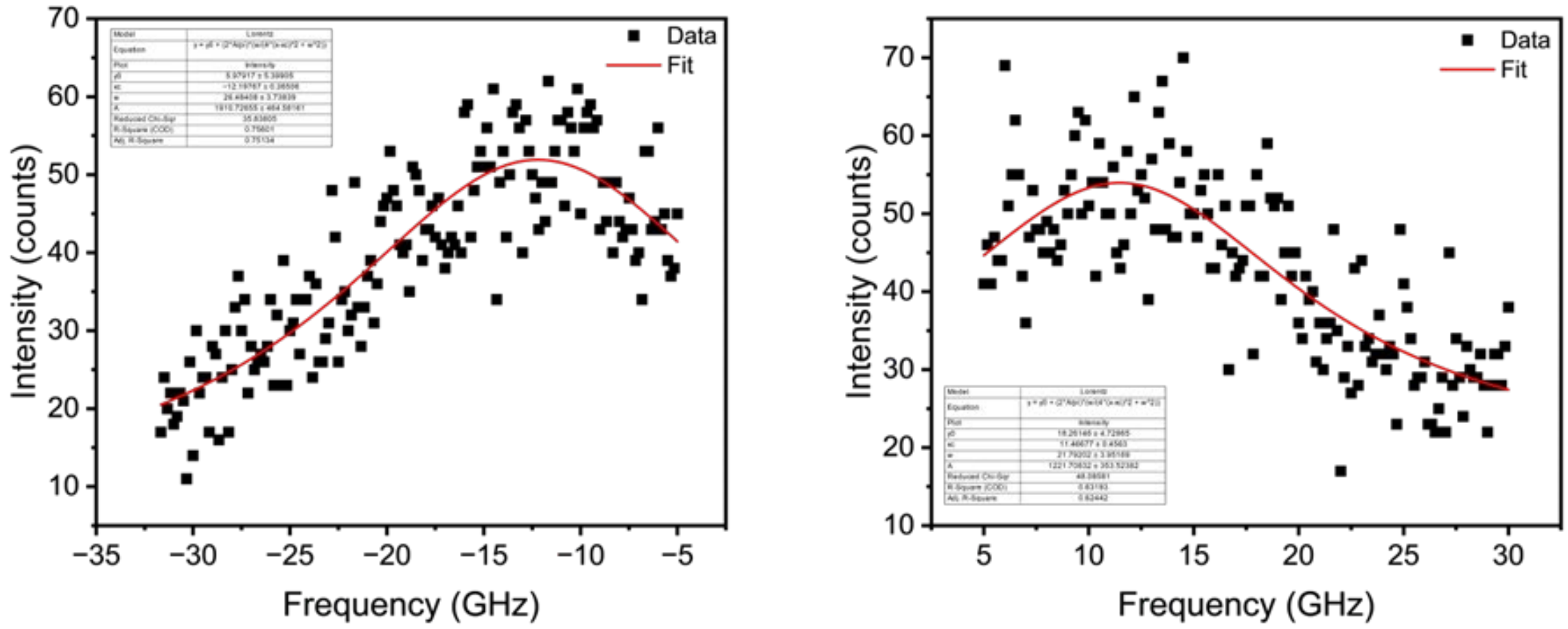

**Figure S-20**: Fit of *Dy*$Mn_6Sn_6$ Stokes (left) and Anti-Stokes (right) magnon at 80 mT.

**Alt text**: Each figure displays raw Brillouin light scattering spectra fitted with a red Lorentzian line shape, plotting intensity against frequency shift to capture the broad Stokes and anti-Stokes magnon features of the dysprosium compound across multiple field values.

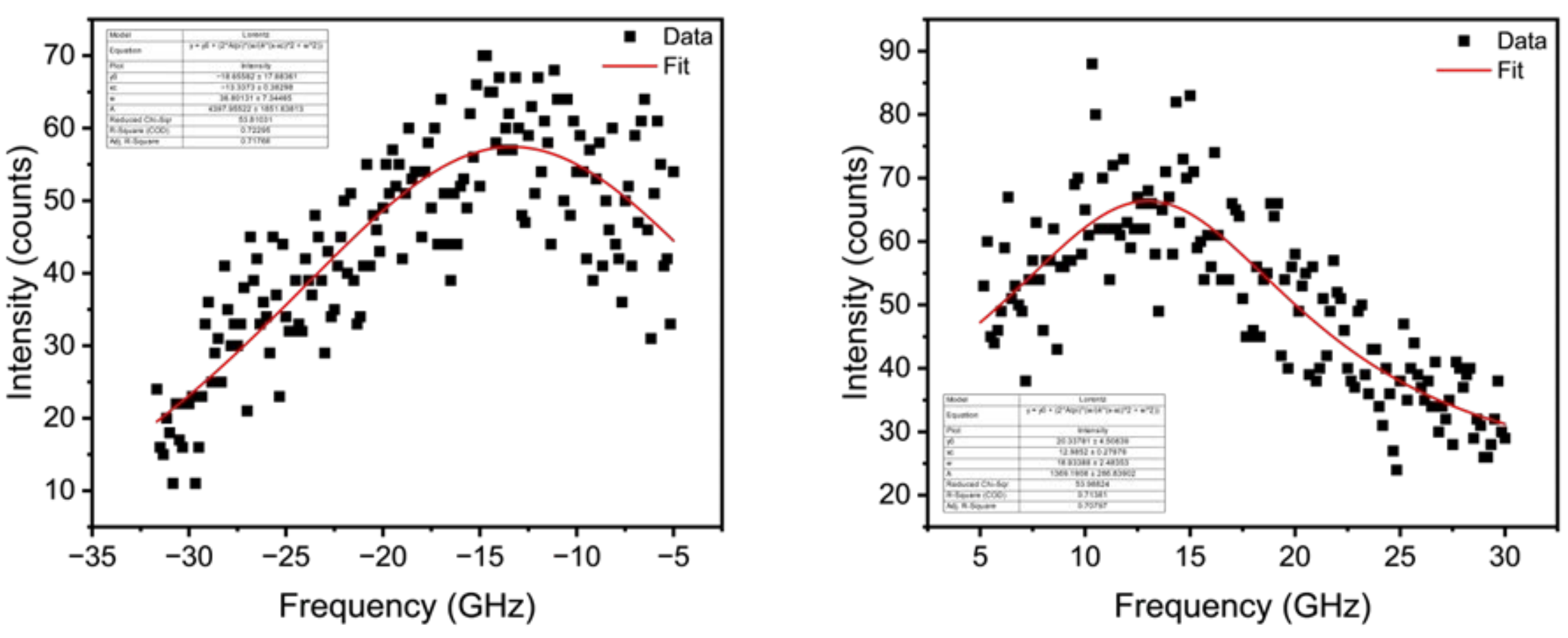

**Figure S-21**: Fit of *Dy*$Mn_6Sn_6$ Stokes (left) and Anti-Stokes (right) magnon at 100 mT.

**Alt text**: Each figure displays raw Brillouin light scattering spectra fitted with a red Lorentzian line shape, plotting intensity against frequency shift to capture the broad Stokes and anti-Stokes magnon features of the dysprosium compound across multiple field values.

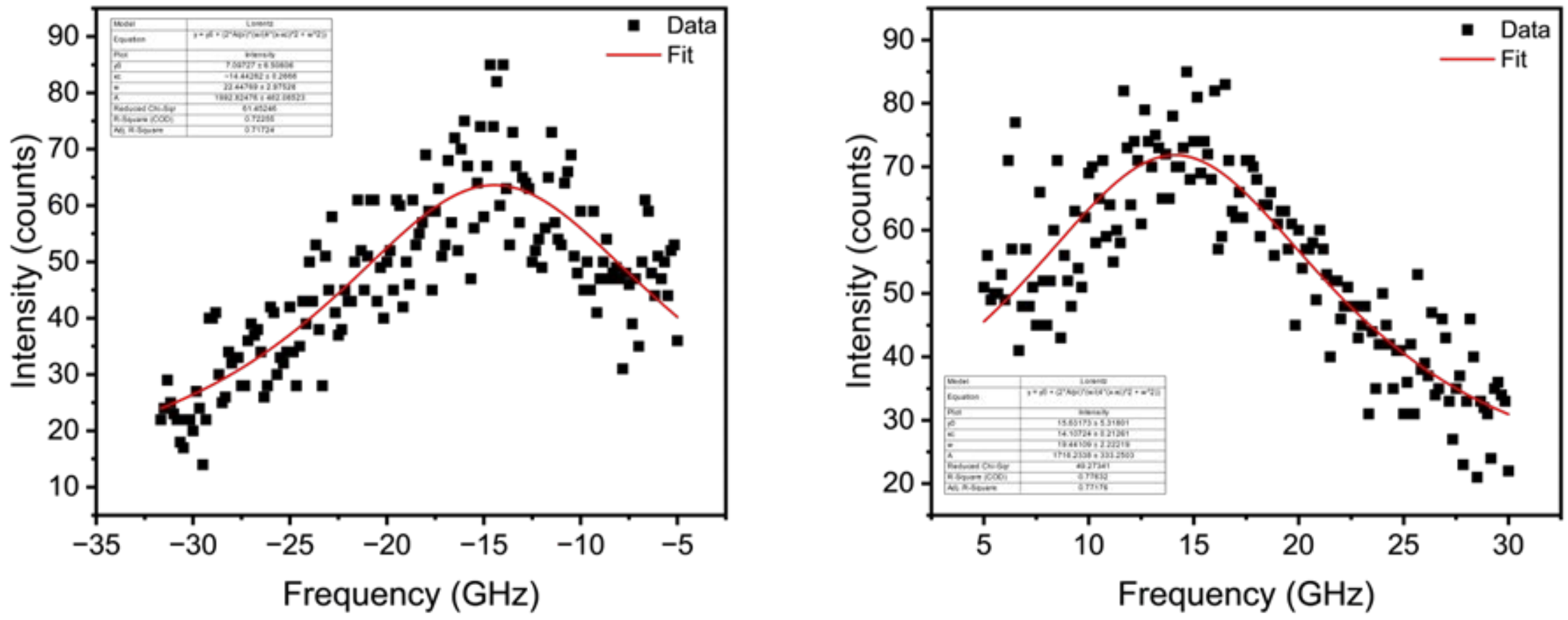

**Figure S-22**: Fit of $DyMn_6Sn_6$ Stokes (left) and Anti-Stokes (right) magnon at 120 mT.

**Alt text**: Each figure displays raw Brillouin light scattering spectra fitted with a red Lorentzian line shape, plotting intensity against frequency shift to capture the broad Stokes and anti-Stokes magnon features of the dysprosium compound across multiple field values.

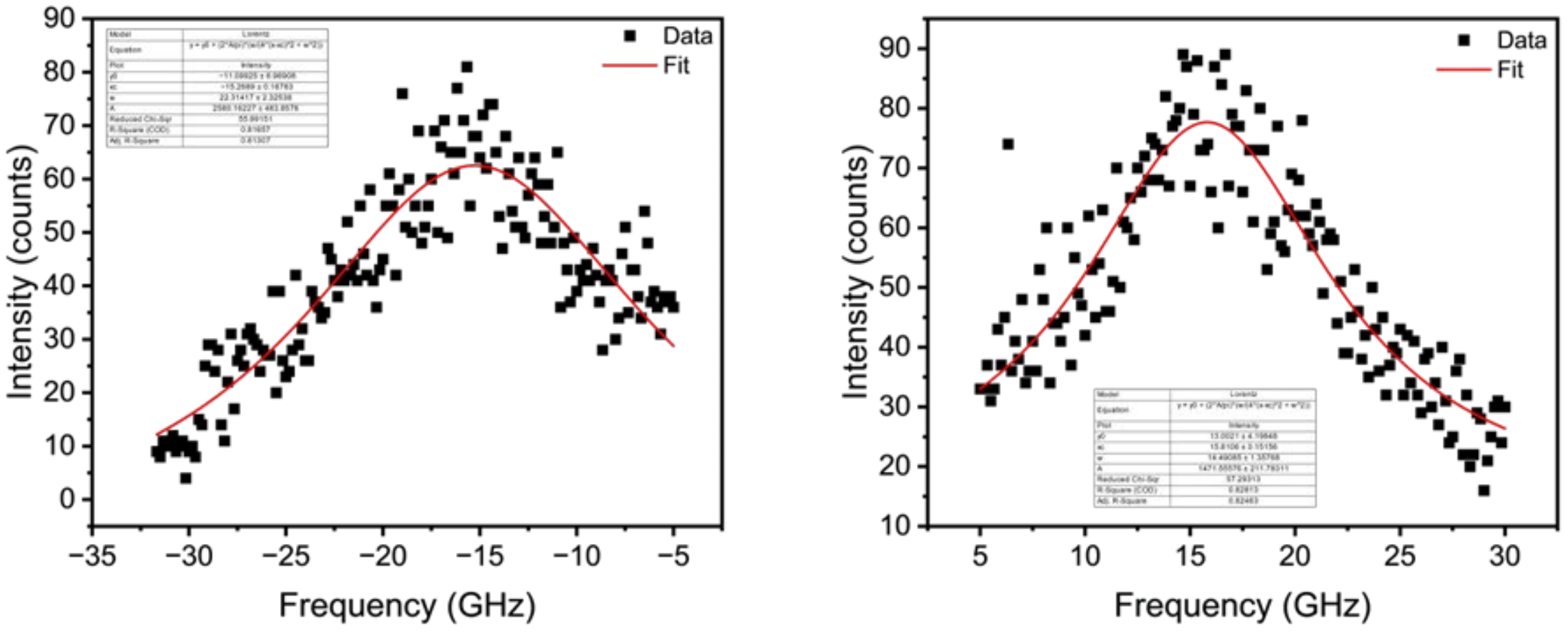

**Figure S-23**: Fit of $DyMn_6Sn_6$ Stokes (left) and Anti-Stokes (right) magnon at 140 mT.

**Alt text**: Each figure displays raw Brillouin light scattering spectra fitted with a red Lorentzian line shape, plotting intensity against frequency shift to capture the broad Stokes and anti-Stokes magnon features of the dysprosium compound across multiple field values.

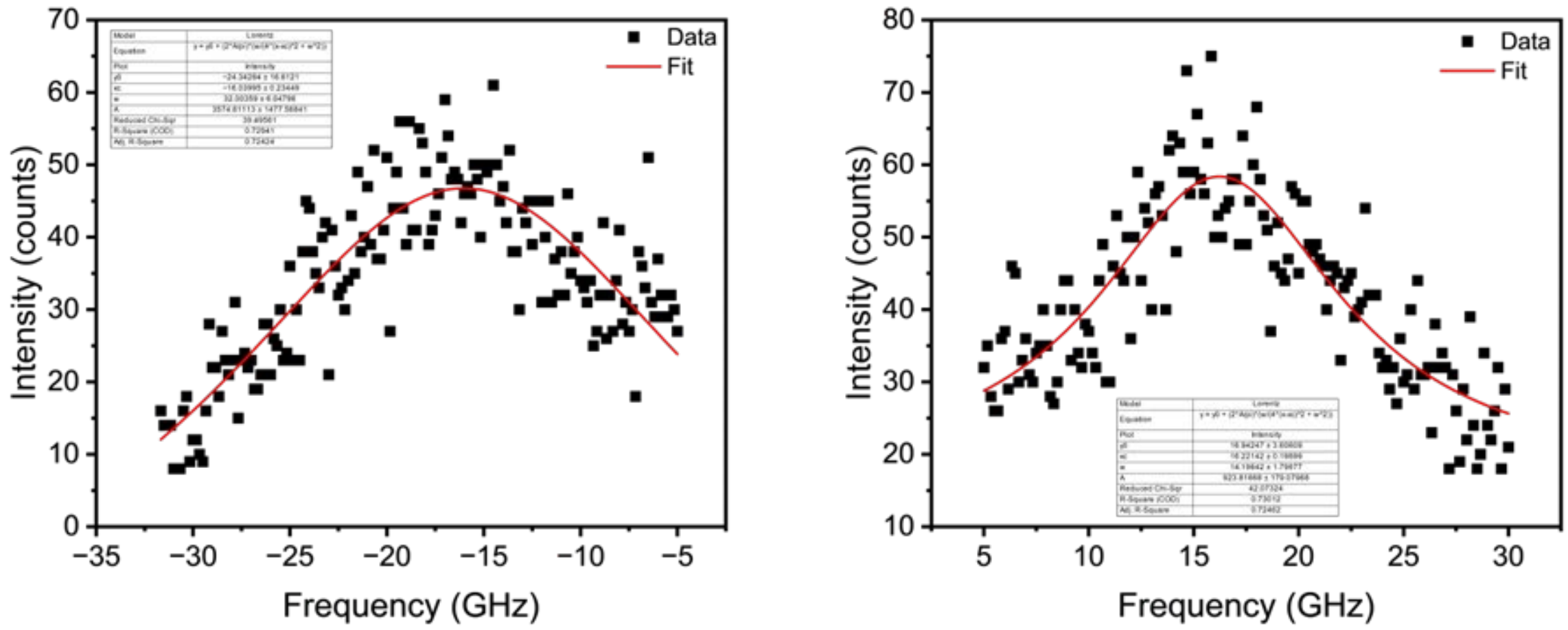

**Figure S-24**: Fit of *Dy*$Mn_6Sn_6$ Stokes (left) and Anti-Stokes (right) magnon at 160 mT.

**Alt text**: Each figure displays raw Brillouin light scattering spectra fitted with a red Lorentzian line shape, plotting intensity against frequency shift to capture the broad Stokes and anti-Stokes magnon features of the dysprosium compound across multiple field values.

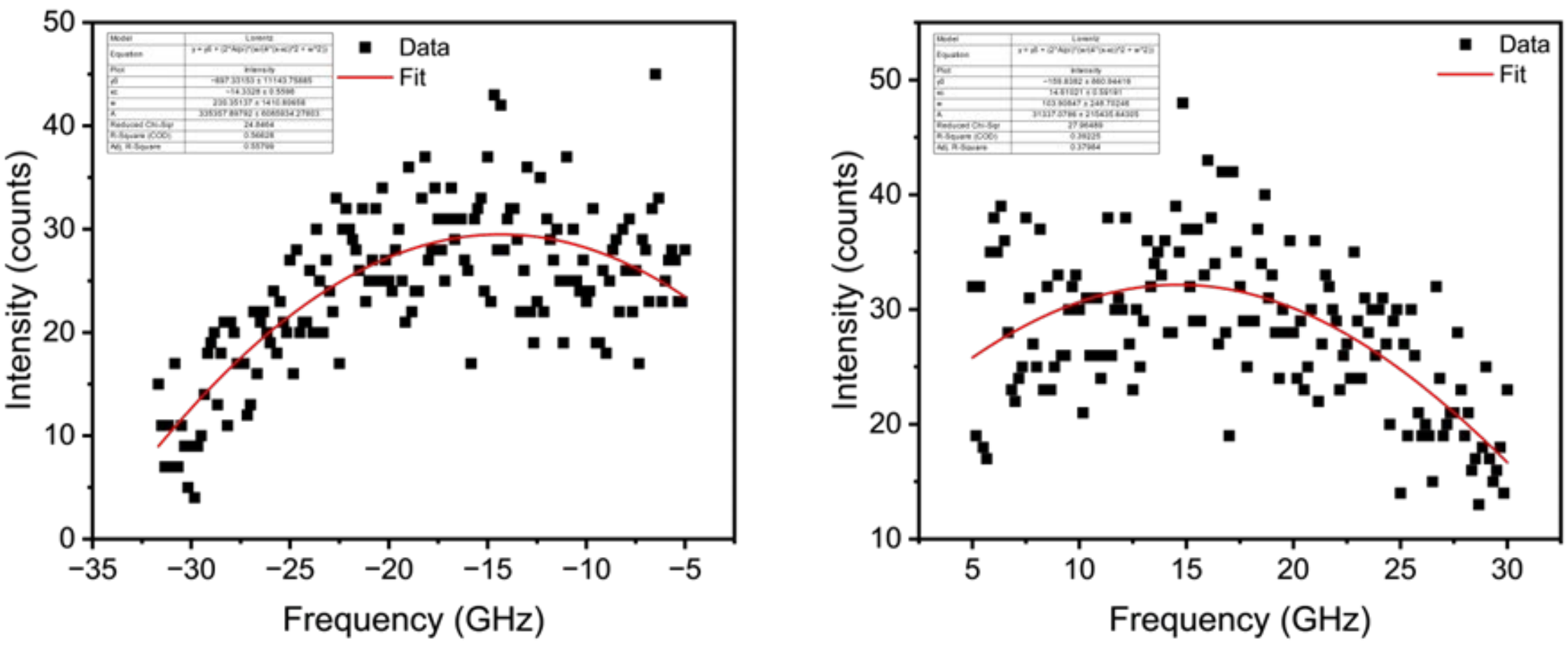

**Figure S-25**: Fit of *Dy*$Mn_6Sn_6$ Stokes (left) and Anti-Stokes (right) magnon at 180 mT.

**Alt text**: Each figure displays raw Brillouin light scattering spectra fitted with a red Lorentzian line shape, plotting intensity against frequency shift to capture the broad Stokes and anti-Stokes magnon features of the dysprosium compound across multiple field values.

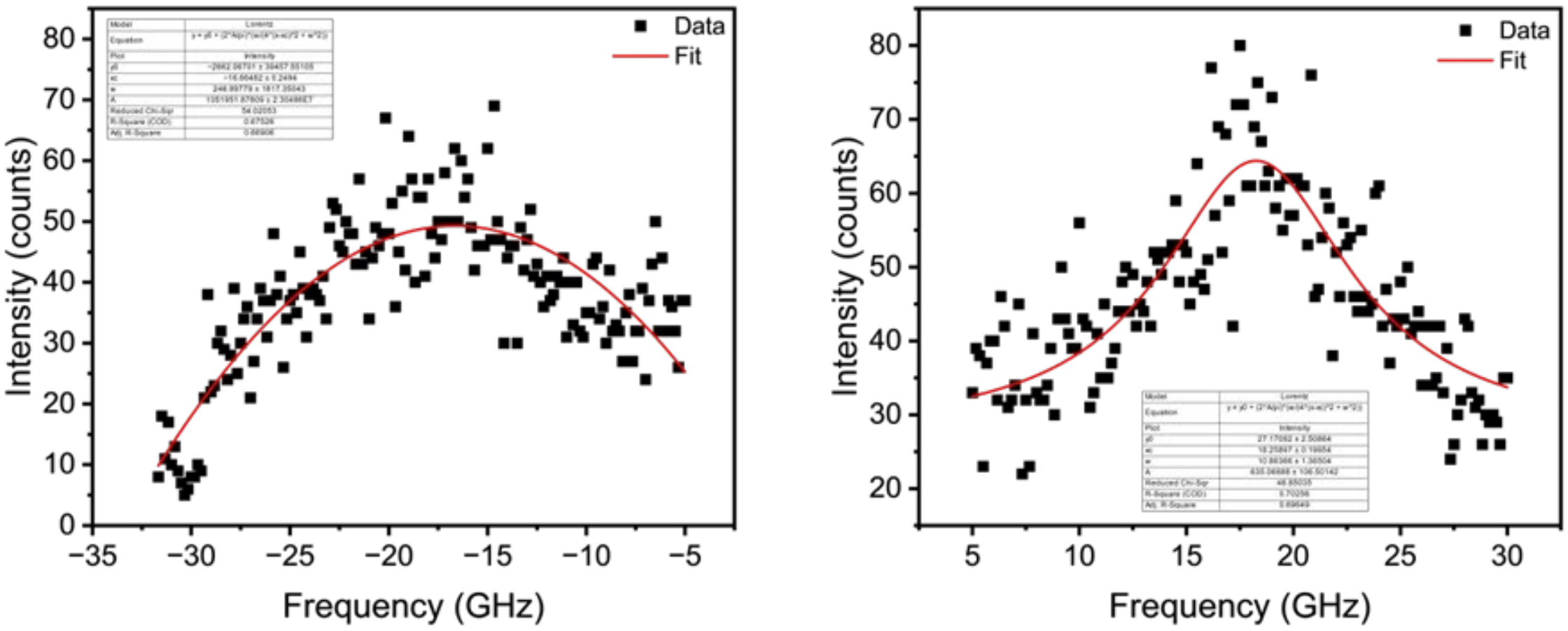

**Figure S-26**: Fit of $DyMn_6Sn_6$ Stokes (left) and Anti-Stokes (right) magnon at 200 mT.

**Alt text**: Each figure displays raw Brillouin light scattering spectra fitted with a red Lorentzian line shape, plotting intensity against frequency shift to capture the broad Stokes and anti-Stokes magnon features of the dysprosium compound across multiple field values.

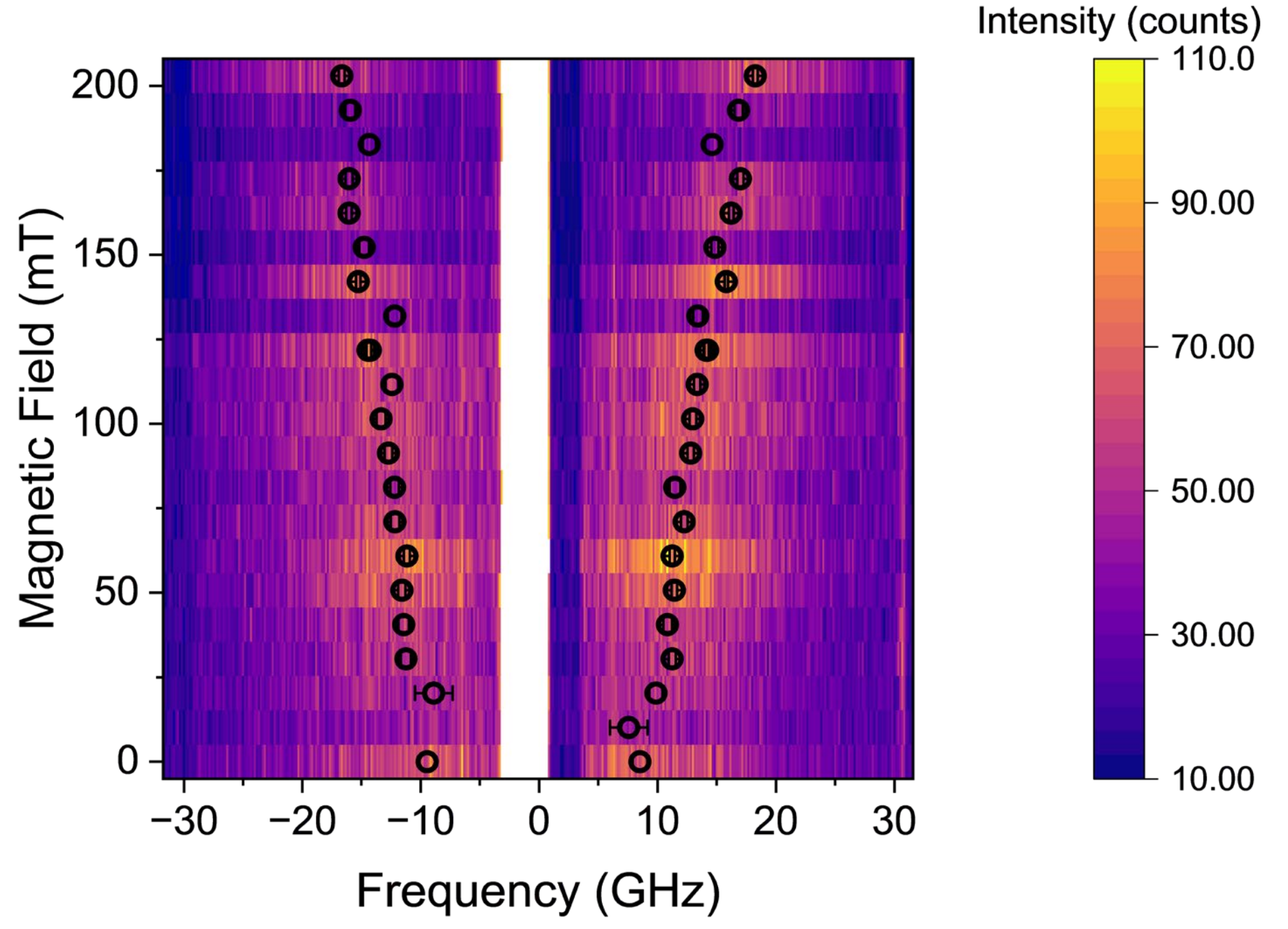

**Figure S-27**: Heatmap of $DyMn_6Sn_6$ with fitted magnon frequencies overlaid as circles.

**Alt text:** A 2D intensity heatmap. The plot displays the field-dependent light scattering intensity profiles for the dysprosium compound across a frequency window, with calculated magnon centers superimposed as circular markers.

# Room temperature BLS spectra fits of 3-Ho

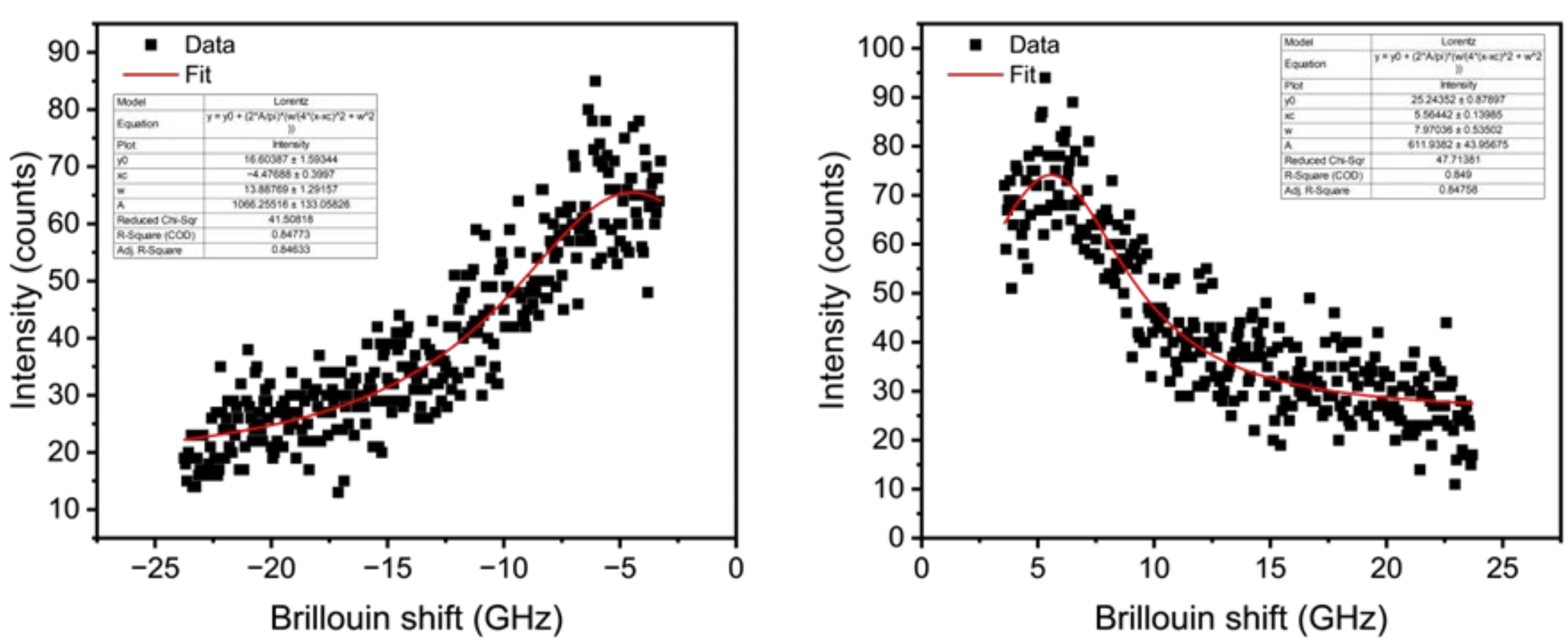

**Figure S-28**: Fit of $HoMn_6Sn_6$ Stokes (left) and Anti-Stokes (right) magnon at 0 mT.

**Alt text**: Each figure shows raw Brillouin light scattering data points fitted with a Lorentzian curve, plotting scattering intensity against frequency shift to capture the low-frequency Stokes and anti-Stokes magnon excitations for the holmium compound at sequential magnetic field values.

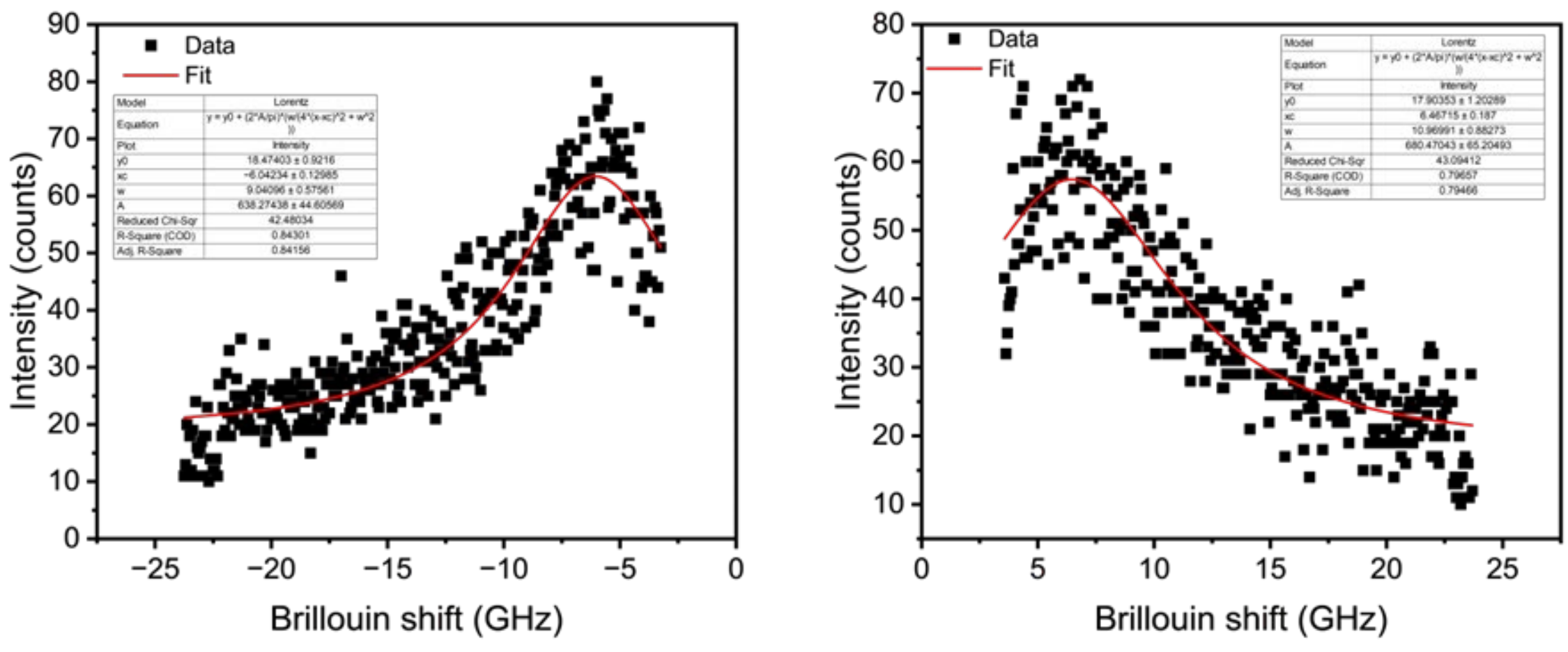

**Figure S-29**: Fit of $HoMn_6Sn_6$ Stokes (left) and Anti-Stokes (right) magnon at 20 mT.

**Alt text**: Each figure shows raw Brillouin light scattering data points fitted with a Lorentzian curve, plotting scattering intensity against frequency shift to capture the low-frequency Stokes and anti-Stokes magnon excitations for the holmium compound at sequential magnetic field values.

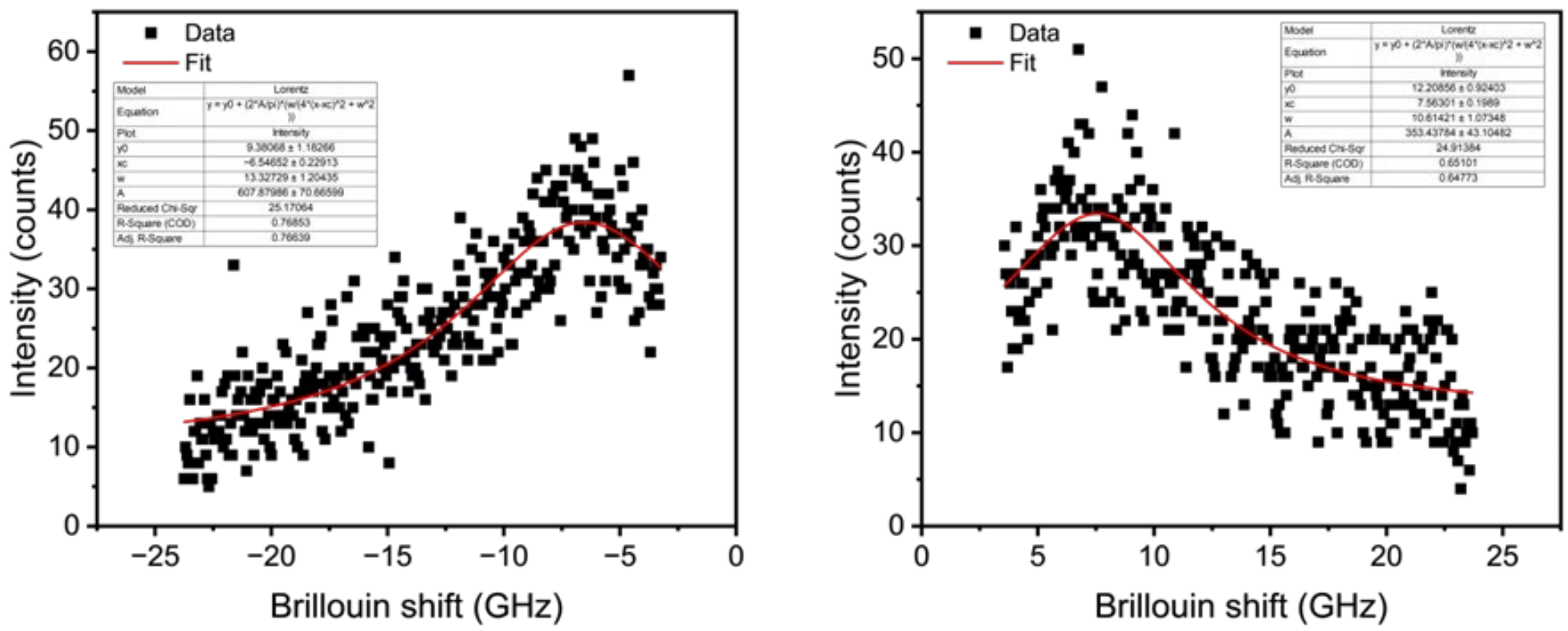

**Figure S-30**: Fit of $HoMn_6Sn_6$ Stokes (left) and Anti-Stokes (right) magnon at 40 mT.

**Alt text**: Each figure shows raw Brillouin light scattering data points fitted with a Lorentzian curve, plotting scattering intensity against frequency shift to capture the low-frequency Stokes and anti-Stokes magnon excitations for the holmium compound at sequential magnetic field values.

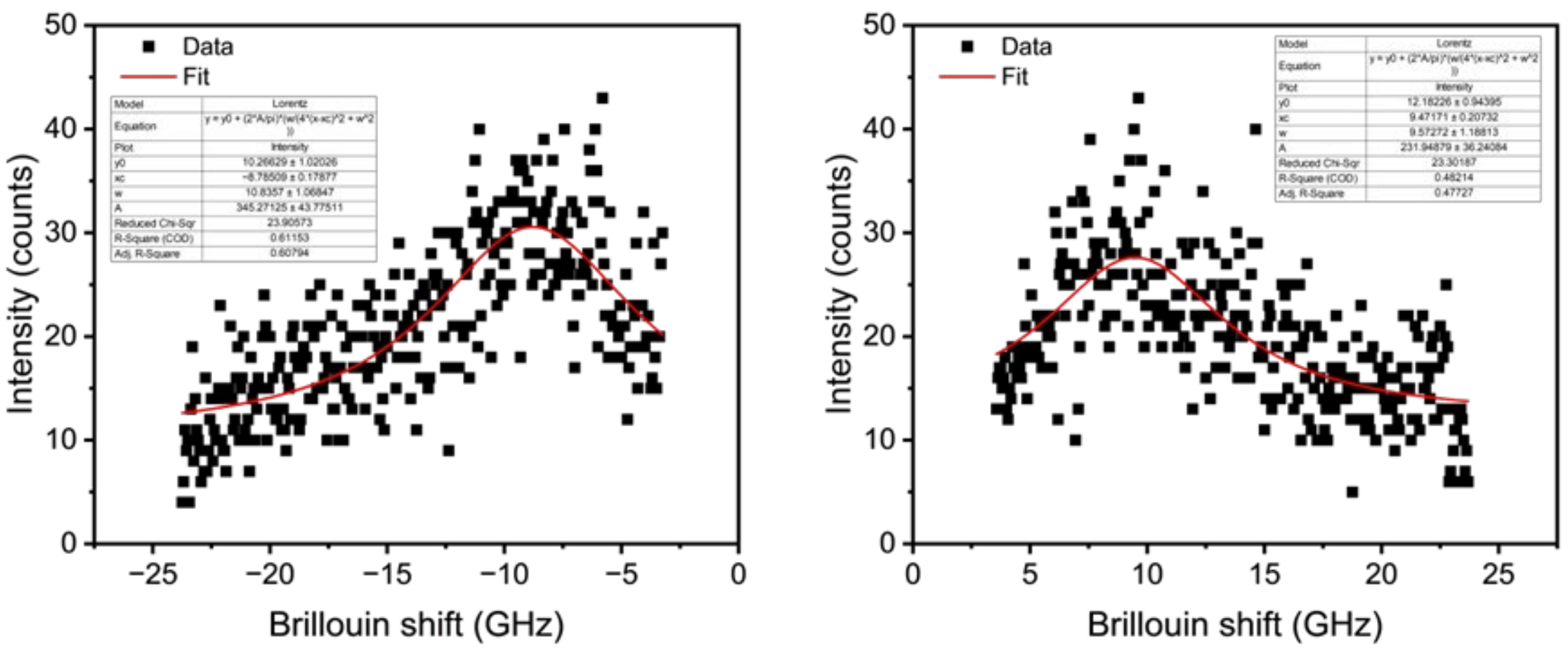

**Figure S-31**: Fit of $HoMn_6Sn_6$ Stokes (left) and Anti-Stokes (right) magnon at 60 mT.

**Alt text**: Each figure shows raw Brillouin light scattering data points fitted with a Lorentzian curve, plotting scattering intensity against frequency shift to capture the low-frequency Stokes and anti-Stokes magnon excitations for the holmium compound at sequential magnetic field values.

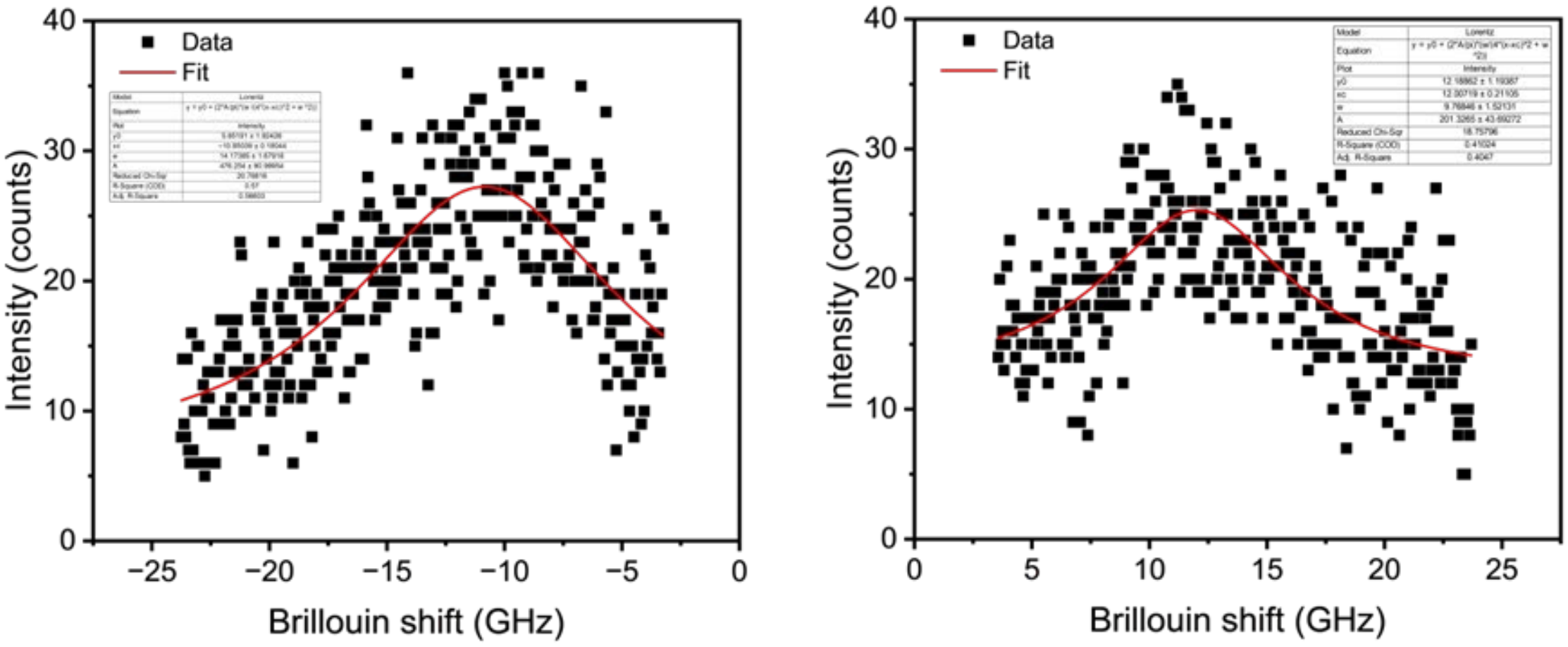

**Figure S-32**: Fit of $HoMn_6Sn_6$ Stokes (left) and Anti-Stokes (right) magnon at 80 mT.

**Alt text**: Each figure shows raw Brillouin light scattering data points fitted with a Lorentzian curve, plotting scattering intensity against frequency shift to capture the low-frequency Stokes and anti-Stokes magnon excitations for the holmium compound at sequential magnetic field values.

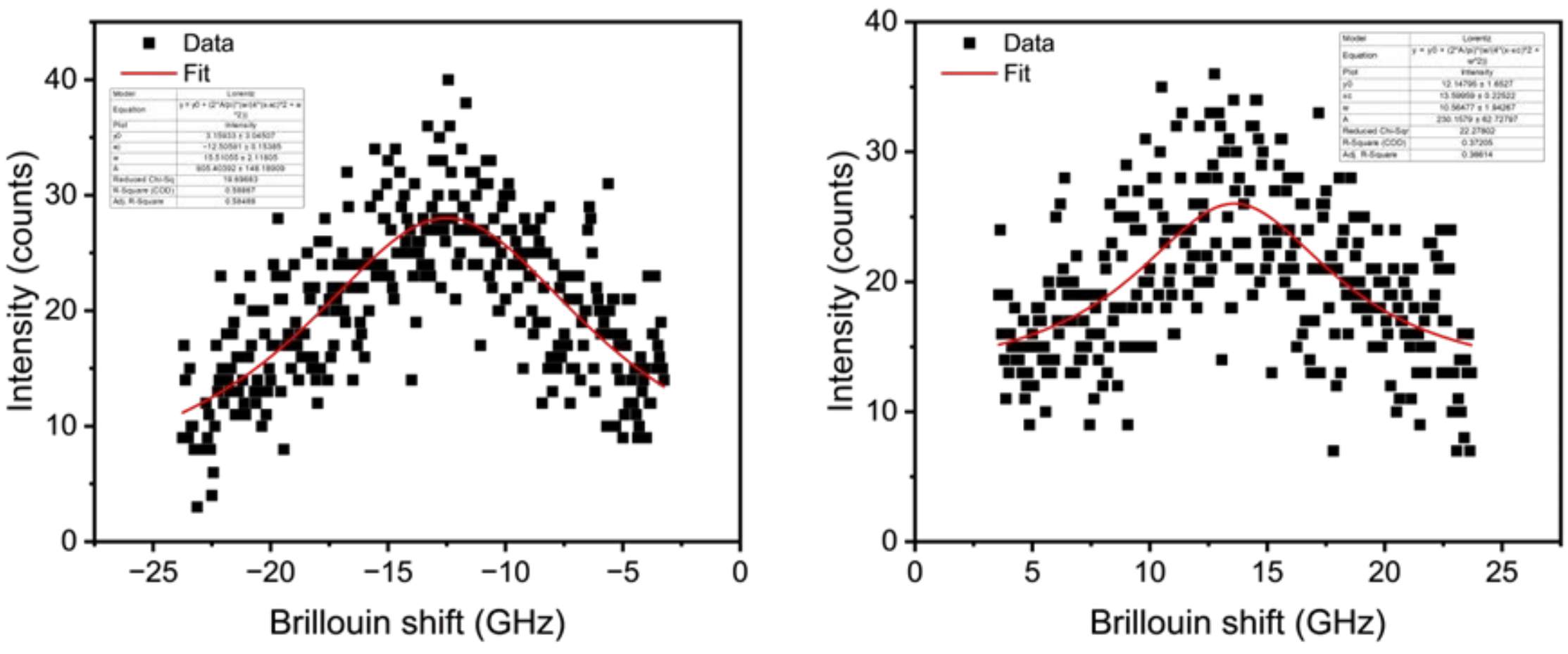

**Figure S-33**: Fit of $HoMn_6Sn_6$ Stokes (left) and Anti-Stokes (right) magnon at 100 mT.

**Alt text**: Each figure shows raw Brillouin light scattering data points fitted with a Lorentzian curve, plotting scattering intensity against frequency shift to capture the low-frequency Stokes and anti-Stokes magnon excitations for the holmium compound at sequential magnetic field values.

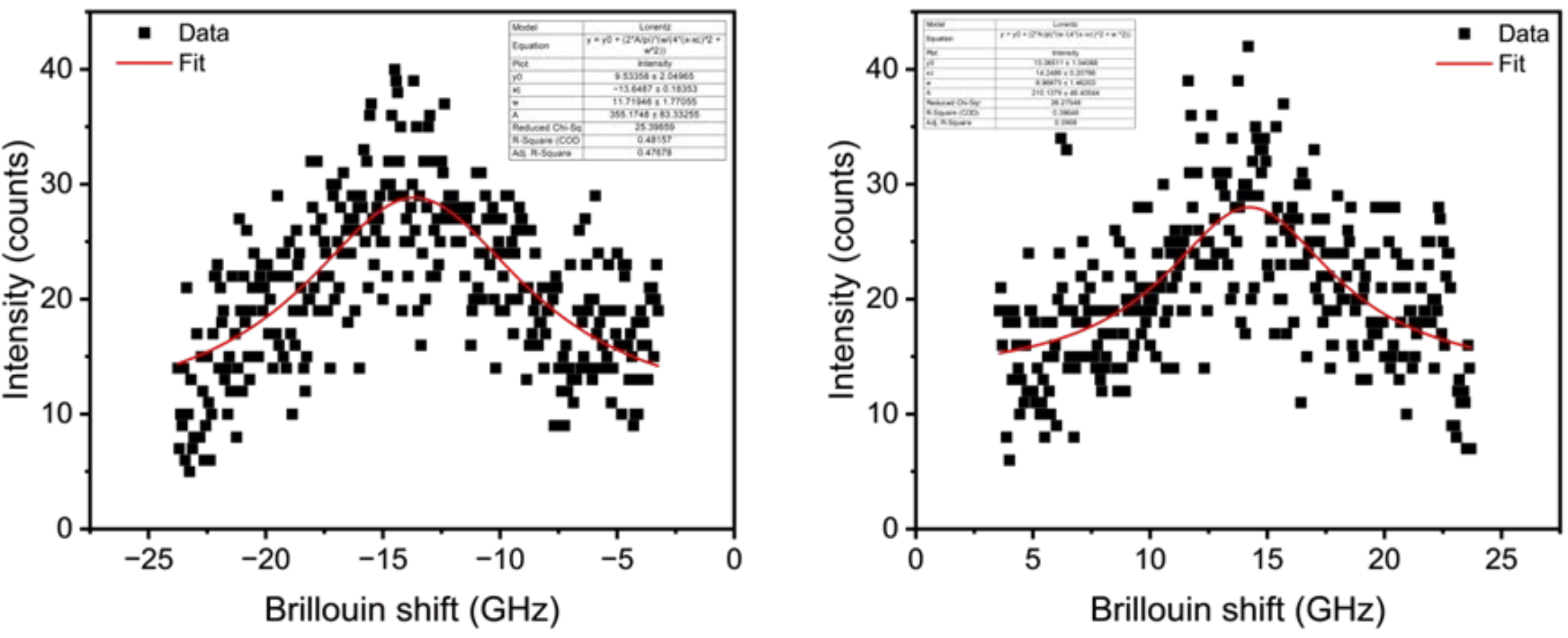

**Figure S-34**: Fit of $HoMn_6Sn_6$ Stokes (left) and Anti-Stokes (right) magnon at 120 mT.

**Alt text**: Each figure shows raw Brillouin light scattering data points fitted with a Lorentzian curve, plotting scattering intensity against frequency shift to capture the low-frequency Stokes and anti-Stokes magnon excitations for the holmium compound at sequential magnetic field values.

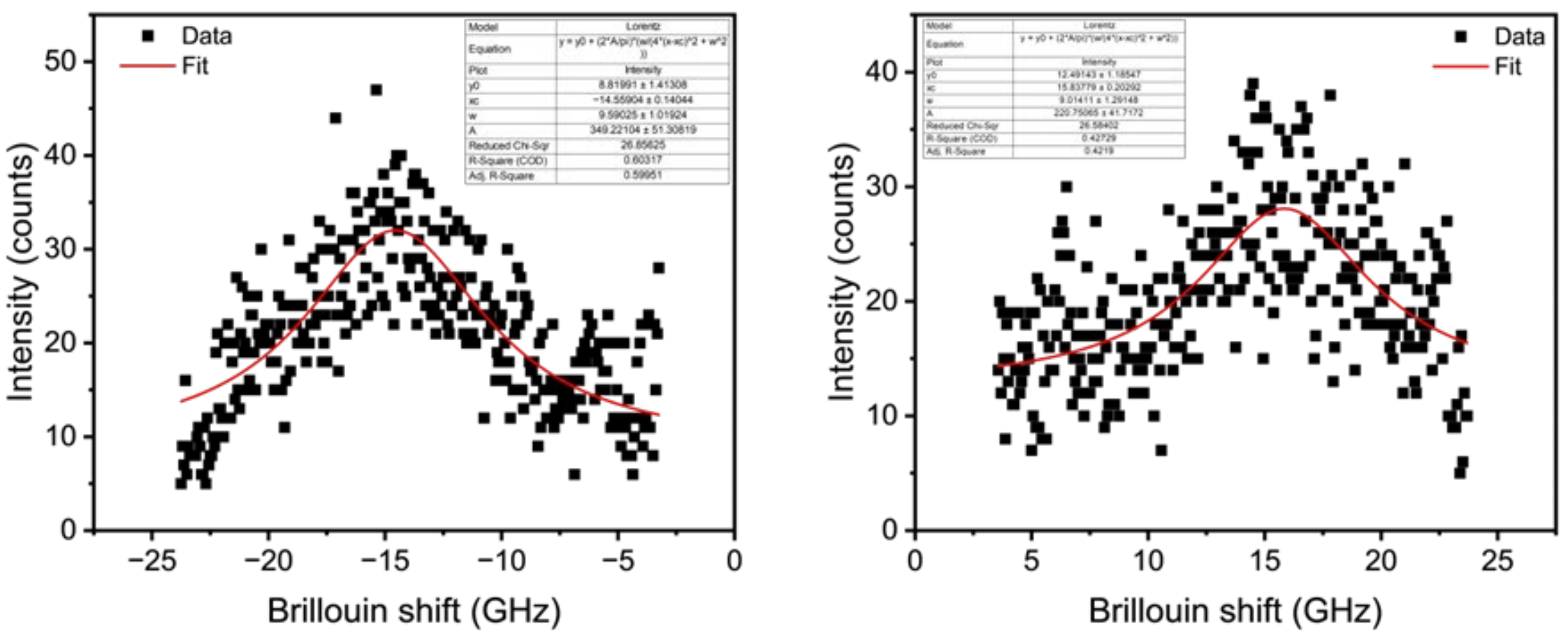

**Figure S-35**: Fit of $HoMn_6Sn_6$ Stokes (left) and Anti-Stokes (right) magnon at 140 mT.

**Alt text**: Each figure shows raw Brillouin light scattering data points fitted with a Lorentzian curve, plotting scattering intensity against frequency shift to capture the low-frequency Stokes and anti-Stokes magnon excitations for the holmium compound at sequential magnetic field values.

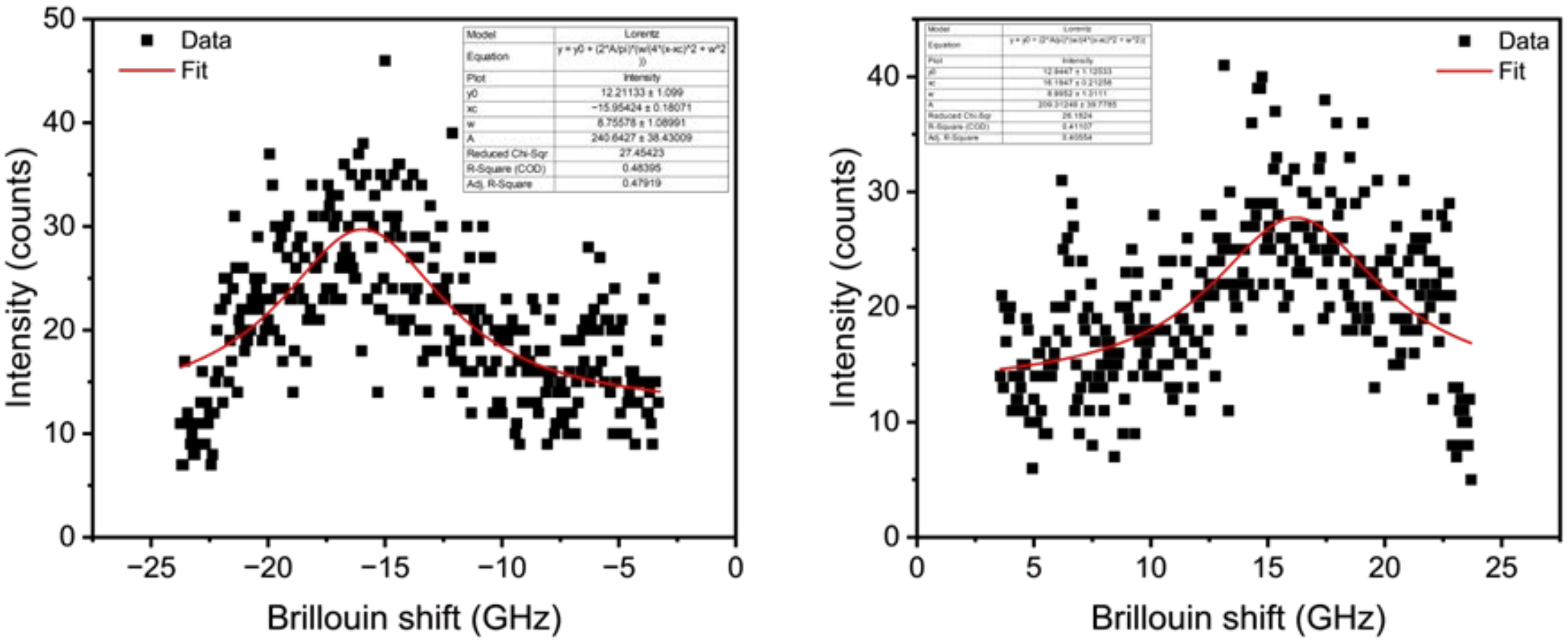

**Figure S-36**: Fit of $HoMn_6Sn_6$ Stokes (left) and Anti-Stokes (right) magnon at 160 mT.

**Alt text**: Each figure shows raw Brillouin light scattering data points fitted with a Lorentzian curve, plotting scattering intensity against frequency shift to capture the low-frequency Stokes and anti-Stokes magnon excitations for the holmium compound at sequential magnetic field values.

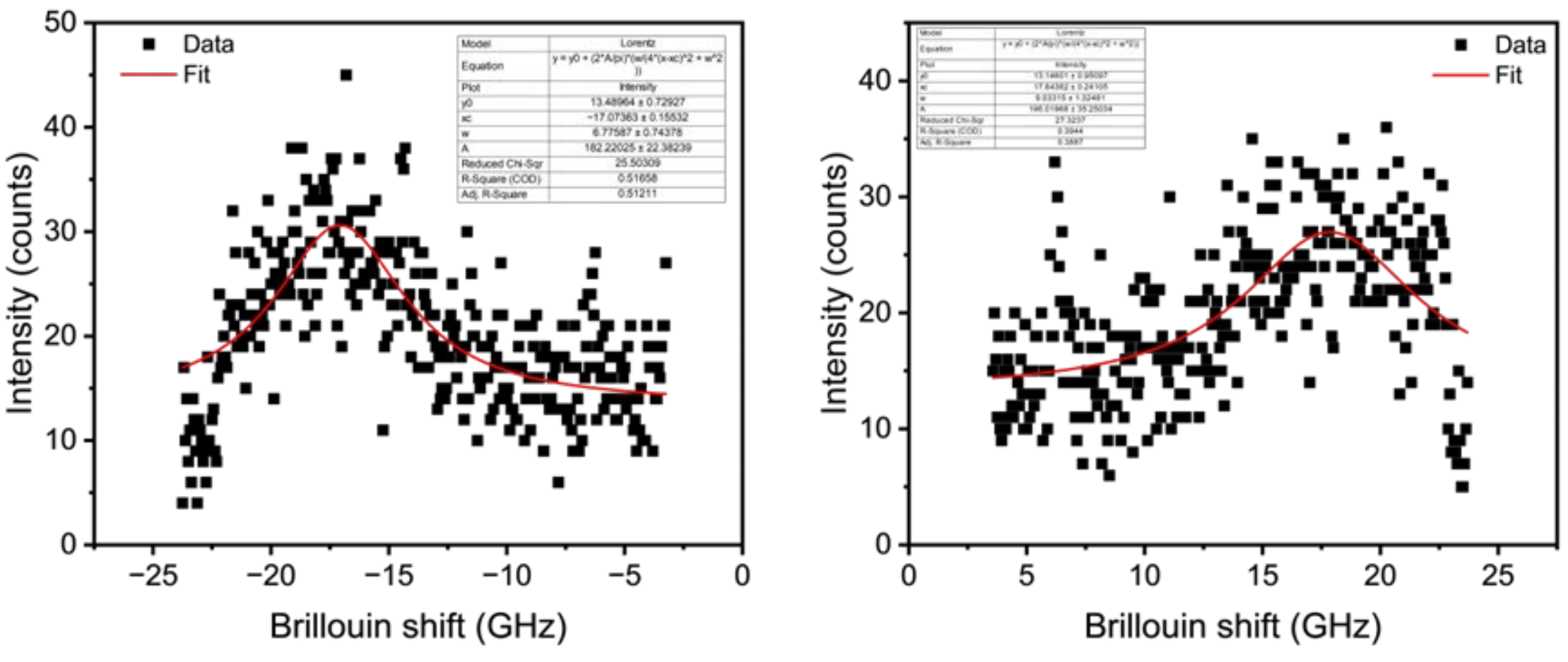

**Figure S-37**: Fit of $HoMn_6Sn_6$ Stokes (left) and Anti-Stokes (right) magnon at 180 mT.

**Alt text**: Each figure shows raw Brillouin light scattering data points fitted with a Lorentzian curve, plotting scattering intensity against frequency shift to capture the low-frequency Stokes and anti-Stokes magnon excitations for the holmium compound at sequential magnetic field values.

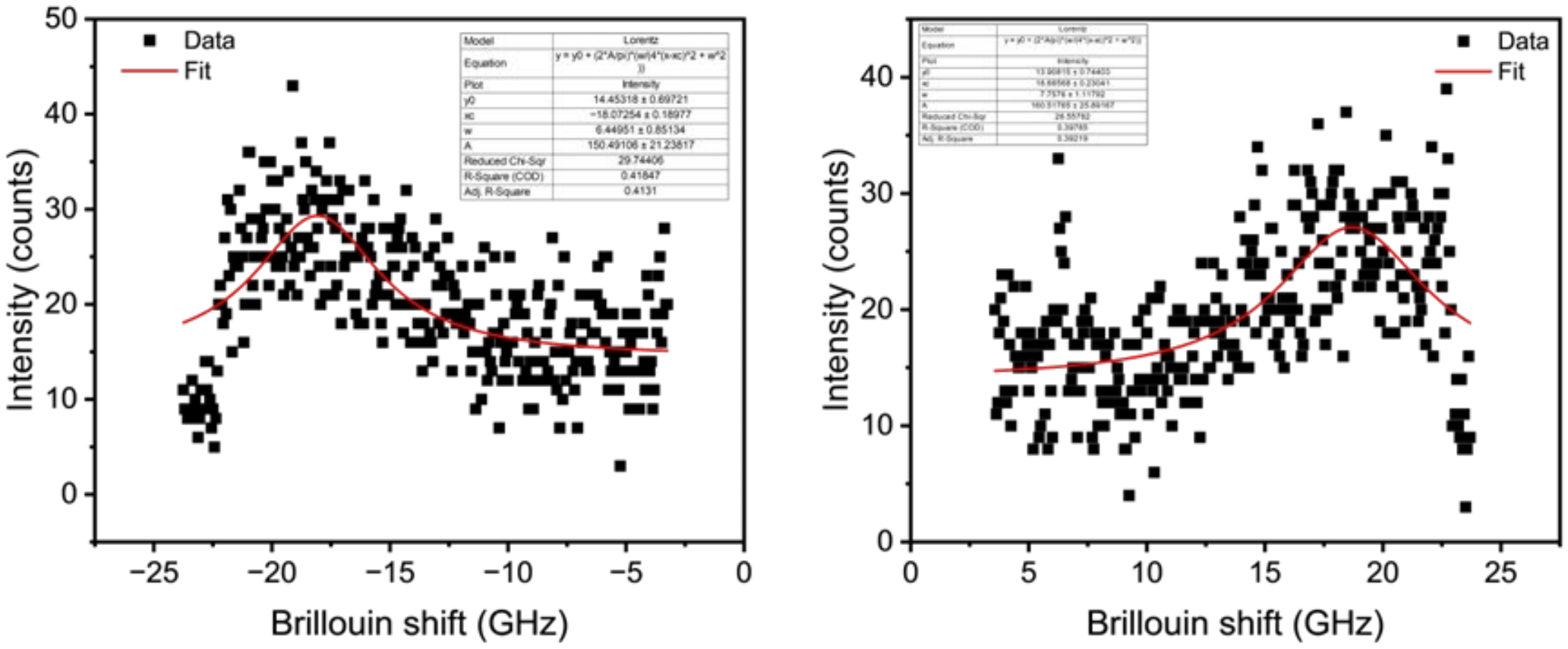

**Figure S-38**: Fit of $HoMn_6Sn_6$ Stokes (left) and Anti-Stokes (right) magnon at 200 mT.

**Alt text**: Each figure shows raw Brillouin light scattering data points fitted with a Lorentzian curve, plotting scattering intensity against frequency shift to capture the low-frequency Stokes and anti-Stokes magnon excitations for the holmium compound at sequential magnetic field values.

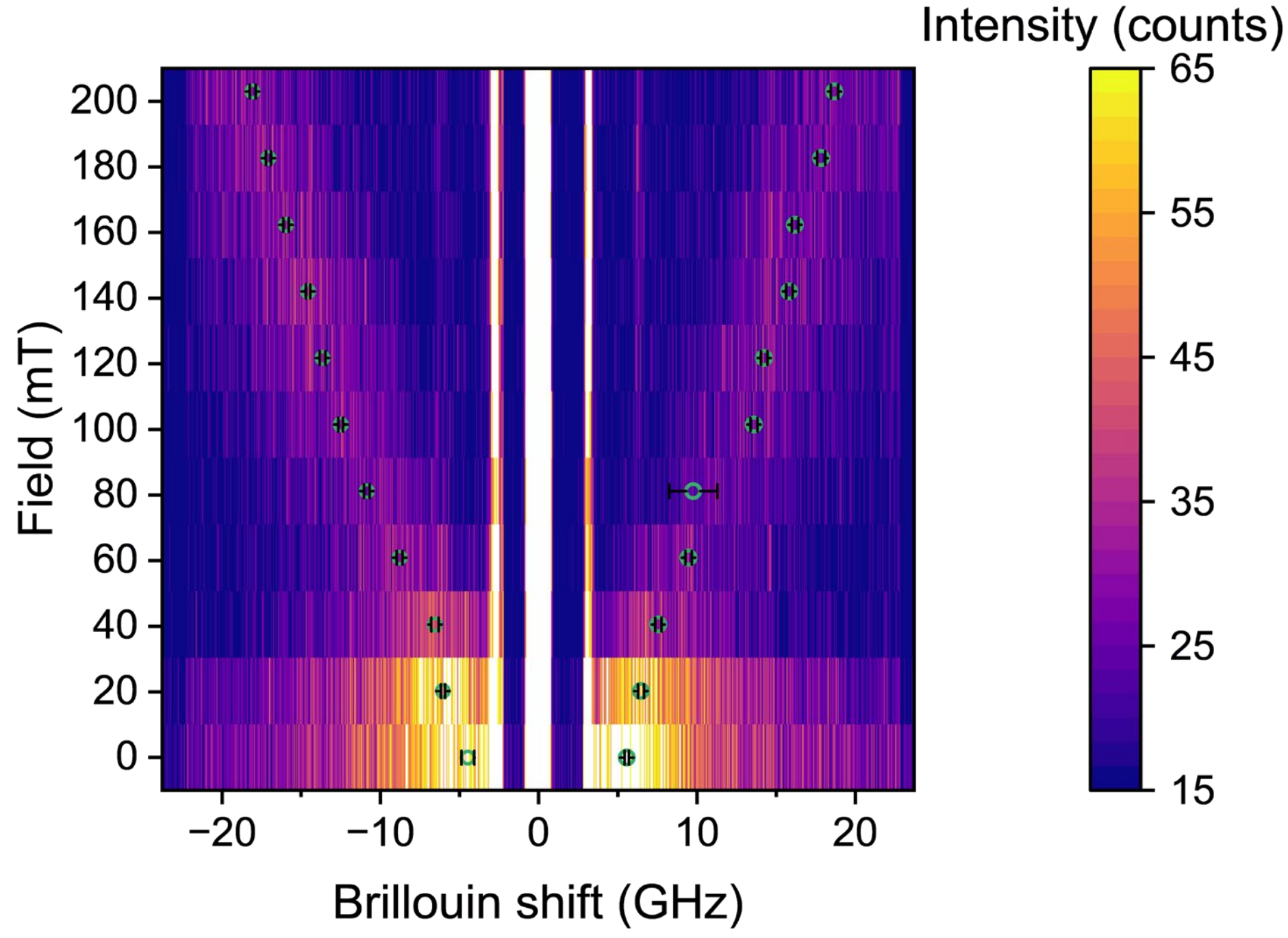

**Figure S-39**: Heatmap of HoMn6Sn6 with fitted magnon frequencies plotted as circles.

**Alt text:** A 2D intensity heatmap. The chart maps the measured light scattering intensity of the holmium compound against applied magnetic field and frequency shift, with fitted magnon frequencies tracked by circular markers.

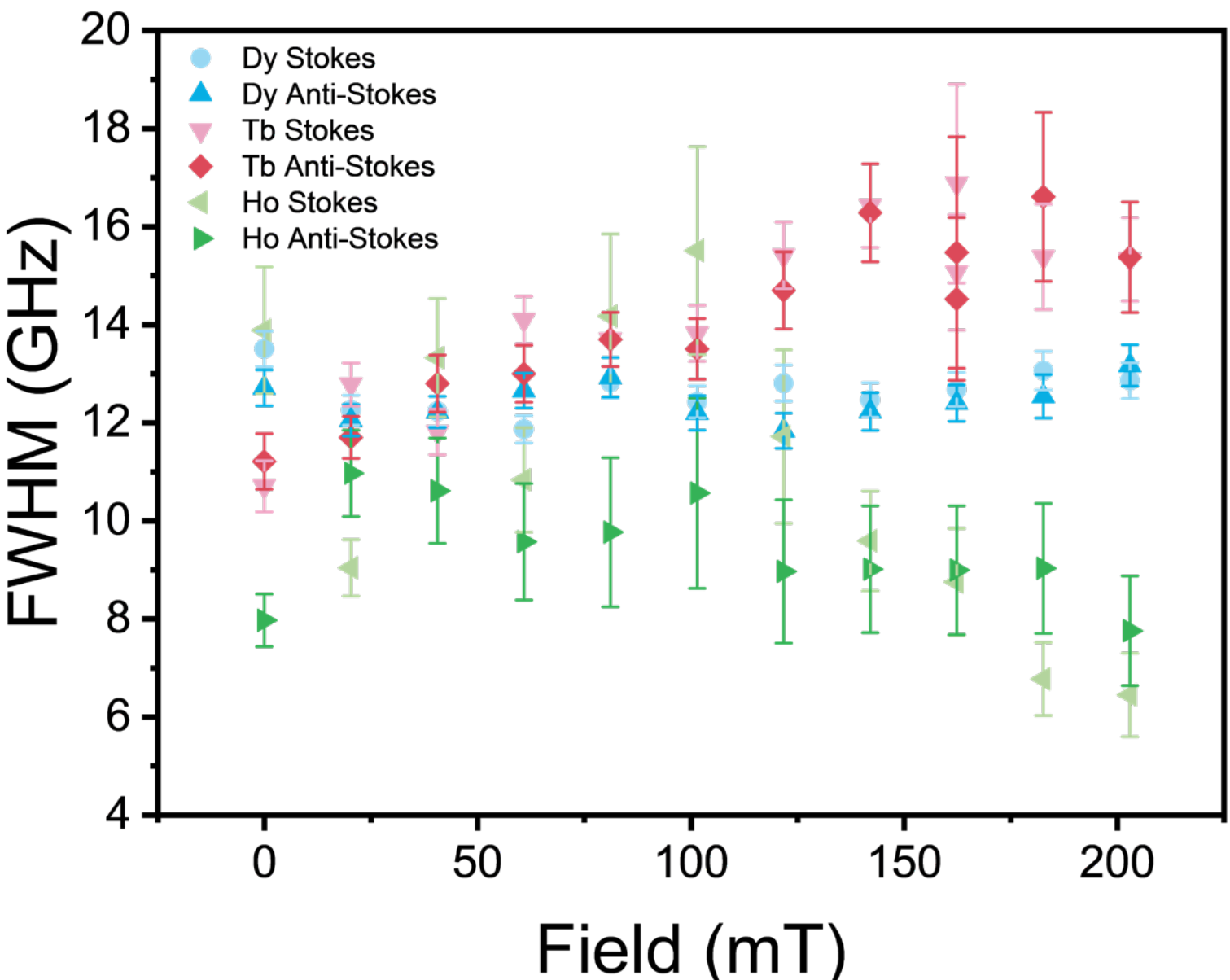

**Figure S-40**: FWHM of **1-Tb**, **2-Dy**, and **3-Ho** magnons as a function of magnetic field.

**Alt text:** A multi-dataset scatter plot. The graph compares the extracted full-width at half-maximum linewidth values against applied magnetic field for the terbium, dysprosium, and holmium magnon peaks, with error bars indicating fit uncertainties.